\documentclass{mnras}
\usepackage{natbib,times}
\usepackage{graphicx}
\usepackage{amsmath}
\usepackage{url}
\usepackage{pdflscape}
\bibpunct[, ]{(}{)}{;}{a}{}{,}

\def \cbat {CBAT}
\def \aj {AJ}
\def \mnras {MNRAS}
\def \pasp {PASP}
\def \apj {ApJ}
\def \apjs {ApJS}
\def \apjl {ApJL}
\def \aap {A\&A}
\def \nat {Nature}
\def \araa {ARAA}

\def \aaps {A\&A Suppl.}

\def \na {New Astronomy}

\newcommand{\kms} {$\mathrm{ km \; s^{-1}}\,$}
\newcommand{\msol} {M$_{\odot}$}
\newcommand{\mza} {M$_{ZAMS}$}

\newcommand{\about} {$\sim$}
\newcommand{\metal}{12 + log(O/H) }

\def\lesssim{\mathrel{\hbox{\rlap{\hbox{\lower4pt\hbox{$\sim$}}}\hbox{$<$}}}}
\def\gtrsim{\mathrel{\hbox{\rlap{\hbox{\lower4pt\hbox{$\sim$}}}\hbox{$>$}}}}
\newcommand{\halpha} {$\mathrm{H\alpha}$ }
\newcommand{\ang} {$\mathrm{\AA}$}
\newcommand{\degree}{$^{\circ}$}
\newcommand{\hbeta} {$\mathrm{H\beta}\,$}
\long\def\symbolfootnote[#1]#2{\begingroup%
\def\thefootnote{\fnsymbol{footnote}}\footnote[#1]{#2}\endgroup} 
\begin{document}
\title [3D shape of Ic-bl SN~2014ad]{The evolution of the 3D shape of the broad-lined type Ic SN~2014ad} 
\author[Stevance et al.]{
\parbox[t]{\textwidth}{\raggedright
H.F. ~Stevance$^{1}$\thanks{f.stevance1@sheffield.ac.uk}, J.R. ~Maund$^{1}$\thanks{Royal Society Research Fellow}, D. ~Baade$^{2}$, P. ~H\"oflich$^{3}$, S. ~Howerton$^{4}$,\\ F. ~Patat$^{2}$, M. ~Rose$^{1}$, J. ~Spyromilio$^{2}$, J.C. ~Wheeler$^{5}$, L. ~Wang$^{6}$}
\vspace*{6pt}\\
$^{1}$ University of Sheffield, Department of Physics and Astronomy, Hounsfield Rd, Sheffield S3 7RH, UK.\\
$^{2}$ European Organisation for Astronomical Research in the Southern Hemisphere, Karl-Schwarzschild-Str. 2, 85748 Garching. b. M\"unchen, Germany\\
$^{3}$ Department of Physics, Florida State University, 315 Keen Building, Tallahassee, FL 32306-4350, USA\\
$^{4}$ Kansas Astronomical Observers, Arkansas City, KS, USA. \\
$^{5}$ Department of Astronomy, University of Texas at Austin, Austin, TX 78712-1205, USA\\ 
$^{6}$ Department of Physics, Texas A\&M University, College Station, TX 77843, USA \\
}
\maketitle
\begin{abstract}
We present optical spectropolarimetry and spectroscopy of the broad-lined Type Ic (Ic-bl) SN~2014ad. Our spectropolarimetric observations cover 7 epochs, from -2 days to 66 days after V-band maximum, and the spectroscopic data were acquired from -2 days to +107 days. The photospheric velocity estimates showed ejecta speeds similar to those of SN~1998bw and other SNe associated with GRBs. The spectropolarimetric data revealed aspherical outer ejecta and a nearly spherical interior. The polarisation associated with O\,{\sc i} $\lambda$7774 and the Ca\,{\sc ii} infrared triplet suggests a clumpy and highly asymmetrical distribution of these two species within the ejecta. Furthermore it was shown that the two line forming regions must have been spatially distinct and  oxygen was found to have higher velocities than calcium. Another oxygen line-forming region was also identified  much closer to the core of the explosion and distributed in a spherical shell. It is difficult to reconcile the geometry of the deeper ejecta with a jet driven explosion, but the high ejecta velocities of SN 2014ad are typical of those observed in SNe Ic-bl with GRBs, and the behaviour of the oxygen and calcium line-forming regions is consistent with fully jet-driven models. The metallicity of the host galaxy of SN~2014ad was also calculated and compared to that of the hosts of other SNe Ic-bl with and without GRBs, but due to the overlap in the two populations no conclusion could be drawn. 
\end{abstract}


\section{Introduction}
\label{intro}

The deaths of most massive stars (\mza $>$ 8 \msol) result in powerful explosions called Core Collapse Supernovae (CCSNe). Different progenitor stars yield different types of CCSNe, which can be divided into multiple categories according to the characteristics of their lightcurves and spectra. Progenitor stars that have shed their outer layers (so called ``stripped envelope") either via strong winds or through binary interactions with a close companion, will produce Type Ib and Type Ic SNe that are characterised by the presence or absence of He in their spectra,  respectively (for reviews see \citealt{filippenko97}, \citealt{crowther07} and \citealt{smartt09}). Broad-Lined Type Ic SNe (SNe Ic-bl),  show extremely wide and blended spectral features compared to ``normal" Type Ic SNe (e.g. SN~1997X, see \citealt{munari98}; SN~1997ef,  see \citealt{iwamoto00}; or SN~2002ap, see \citealt{mazzali02}), as the result of significantly greater ejecta velocities. A study of 17 Type Ic SNe and 21 SNe Ic-bl conducted by \cite{modjaz16} has shown that the mean peak expansion velocity is \about 10,000~\kms faster in Ic-bl than in Type Ic SNe over all epochs.

SNe Ic-bl have been of particular interest over the past two decades, owing to their relation to long Gamma-Ray Bursts (GRBs) and X-Ray Flashes (XRFs). SN~1998bw \citep{patat01} was the first convincing candidate for the association of a SN~to a long GRB. Since then, many other examples have been studied with various GRB energies -- e.g. SN~2003dh / GRB030329, \citep{stanek03}; SN~2003lw / GRB031203, \citep{malesani04}; SN~2006aj / XRF060218, \citep{sollerman06}; SN~2009nz / GRB091127, \citep{berger11}; SN~2010bh / GRB100316D, \citep{chornock10}; SN~2012bz / GRB120422A, \citep{schulze14}; SN~2013dx / GRB130702A, \citep{delia15}. Not all SNe Ic-bl are part of a SN/GRB pair, however, which, as stated by \cite{soderberg06}, cannot be explained solely through viewing angle effects. This implies that the progenitors and/or explosion mechanisms giving rise to GRB/SNe and GRB-less SNe Ic-bl are distinct in some way. 

The main model proposed to produce GRBs is the Collapsar model, whereby the core of the massive star collapses to a black hole. Accretion onto the black hole taps the rotational energy of the star via magnetic coupling, resulting in collimated jets  which power the explosion and yield the GRB (e.g. \citealt{woosley93,woosleymacfadyen99}). Recent studies have shown that for central engines of short enough lifetime, the jets may fail to break out of the envelope of the progenitor, but would impart sufficient energy to drive a rapid expansion, hence yielding SNe Ic-bl  (\citealt{bromberg11}, \citealt{lazzati12}). The high-pressure cocoon resulting from a jet travelling through stellar material produces a conical shock resulting ejecta that are not spherically symmetric \citep{lazzati12}. 

In spatially unresolved SNe, clues as to the geometry of their ejecta can be inferred from spectropolarimetric observations. The main source of opacity in the atmosphere of a SN at early times is electron scattering, such that the light emitted is linearly polarised \citep{ss82} and the orientation of the polarisation is perpendicular to the plane of scattering. Consequently, in the case of a spatially unresolved spherical photosphere the polarisation components cancel out completely. A departure from spherical symmetry, however, will result in incomplete cancellation and a net non-zero degree of polarisation \citep{ss82, mccall84}. Similarly, if the photosphere is being partially obscured by a given line-forming region, incomplete cancellation will result at the wavelengths corresponding to the absorption features associated with the species present in that line-forming region. Therefore, for spheroidal atmospheres the continuum polarisation is closely related to the axis ratio of the photosphere (e.g. \citealt{hoflich91} and Figure 4 therein), and the polarisation associated with spectral features probes smaller scale and composition related asymmetries \citep{ww08}. 

Virtually all CCSNe exhibit significant departure from spherical symmetry \citep{ww08}, with Type II SNe often showing lower levels of continuum polarisation at early times (\about 0.0 - 0.2\%) increasing at later epochs (\about 0.5 \%; e.g. SN~1999em, \citealt{leonard01} and SN~2004dj, \citealt{leonard06}) whereas stripped envelope SNe and GRB-SNe tend to exhibit higher levels of polarisation at practically all epochs e.g. SN~2008D (\about 0.4\%), SN~2008aq (0.7 - 1.2\%), SN~2006aj (\about 0.5\%), SN~1997X ($>$ 4\%), SN~2003dh (ranging from \about 0.5 to 2\%), SN~1998bw (\about 0.4\% - 0.8\%) (\citealt{maund08D,stevance16,maund06aj,wang01,kawabata03,patat01}) .

\begin{figure}
\centering
\includegraphics[width=8cm]{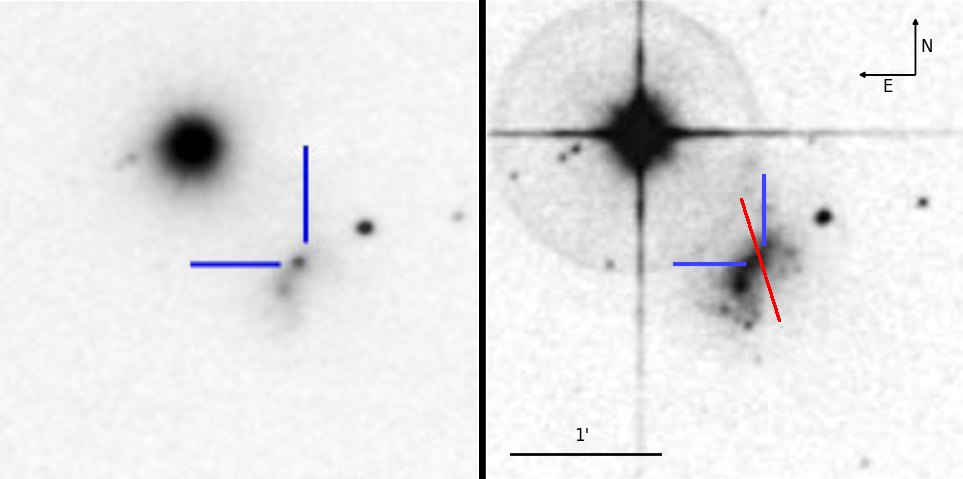}
\caption{Images of MRK 1309 with (left panel) and without (right panel) SN~2014ad. The image on the right hand side is a SERC image from 03rd August 1989  retrieved via Aladin. The image of SN~2014ad in MRK 1309 was obtained from eleven unfiltered 20 second exposures acquired on May 3.1 2014 using a Celestron 11-inch CST and a Orion StarShoot Deep Space Monochrome Imager III. The red line superposed onto the SERC image is the direction of the interstellar polarisation (ISP) determined for MRK 1309 (see section \ref{sec:isp}).  }
\label{fig:14ad}
\end{figure}

In this paper we present the best spectropolarimetric data set obtained as of yet for a SN Ic-bl, extending over 7 epochs, ranging from -2 days to + 66 days with respect to V-band maximum, as well as 8 epochs of spectroscopy ranging from -2 days to +107 days. In Section 2 we give an account of our observations and data reduction; in Section 3 we present our spectroscopic and spectropolarimetric results and in Section 4 we discuss interstellar polarisation and analyse the intrinsic polarisation of SN~2014ad using $q-u$ plots and synthetic V-band polarimetry. Our results and analysis are then discussed in Section 5 and conclusions are given in Section 6.

\section{Observations and Data Reduction}
\label{sec:obs}

\begin{table}
\centering
\label{tab:obs}
\caption{ VLT Observations of SN~2014ad.  The epochs are given relative to the estimated V-band maximum. }
\begin{tabular}{c c c c c}
\hline\hline
Object & Date & Exp. Time & Epoch & Airmass \\
 & (UT) & (s) & (days) & (Avg.)\\
\hline
\hline
\multicolumn{5}{c}{Linear Spectropolarimetry} \\
\hline
SN~2014ad & 2014 March 22 & 16 $\times$ 310 & -2 & 1.069 \\
CD-32d9927\footnotemark[1]  & 2014 March 22 & 2$\times$10 & - & 1.394\\
 \\ 
SN~2014ad & 2014 March 29 & 16 $\times$ 345 & +5 & 1.057 \\
LTT4816\footnotemark[1] & 2014 March 29 & 95 & - & 1.15 \\
 \\
SN~2014ad & 2014 April 11 & 12 $\times$ 345 & +18 & 1.037 \\
LTT4816\footnotemark[1] & 2014 April 11 & 95 & - & 1.117 \\
 \\
SN~2014ad & 2014 April 26 & 8 $\times$ 855 & +33 & 1.072 \\
LTT4816\footnotemark[1]  & 2014 April 26 & 95 & - & 1.109 \\
 \\
SN~2014ad & 2014 May 06 & 8 $\times$ 855 & +43 & 1.061 \\
LTT4816\footnotemark[1]  & 2014 May 06 & 95 & - & 1.41 \\
  \\
SN~2014ad & 2014 May 21 & 8 $\times$ 855 & +58 & 1.080 \\
LTT4816\footnotemark[1]  & 2014 May 21 & 95 & - & 1.159 \\
 \\
SN~2014ad & 2014 May 29 & 8 $\times$ 855 & +66 & 1.092 \\
LTT4816\footnotemark[1]  & 2014 May 29 & 95 & - & 1.172 \\
\hline
\multicolumn{5}{c}{Circular Spectropolarimetry}  \\
\hline
SN~2014ad & 2014 March 29 & 2 $\times$ 345 & +5 & 1.057 \\
LTT4816\footnotemark[1] & 2014 March 29 & 95 & - & 1.15 \\
 \\
SN~2014ad & 2014 April 11 & 4$\times$ 345 & +18 & 1.037 \\
LTT4816\footnotemark[1] & 2014 April 11 & 95 & - & 1.117 \\
\hline
\multicolumn{5}{c}{Spectroscopy} \\
\hline
SN~2014ad & 2014 July 09 & 3 $\times$ 1180 & +107 & 1.38 \\
LTT4816\footnotemark[1] & 2014 July 10 & 2 $\times$ 40 & - & 1.38\\
\hline\hline
\end{tabular}
\end{table}
\footnotetext[1]{Flux Standard}

SN~2014ad was discovered by \citet{14ad} on 12.4 March 2014 in public images from the Catalina Sky Survey (CSS, \citealt{CRTS}) at RA = 11:57:44.44 and $\delta$ = \mbox{-10}:10:15.7. It is located in MRK 1309, with a recessional velocity of 1716~\kms \citep{HIPASS06}. No GRB associated with SN~2014ad was reported. Spectropolarimetric observations of SN~2014ad were conducted with the Very Large Telescope (VLT) of the European Southern Observatory (ESO) using the Focal Reducer and low-dispersion Spectrograph (FORS2) in its  dual-beam spectropolarimeter ``PMOS" mode \citep{1998Msngr..94....1A}. The order sorting filter GG435 was used, but made all wavelengths below 4450 \r{AA} inaccessible to us. Linear spectropolarimetric data of SN~2014ad were obtained for 4 half-wave retarder plate angles (0\degree, 22.5\degree, 45\degree, 67.5\degree) at 7 epochs between 22 March 2014 and 29 May 2014 . Circular spectropolarimetric data were also acquired on 29 March 2014 and 11 April 2014, at quarter wave retarder plate angles -45\degree and 45\degree. Additionally, spectroscopic data was taken on July 2014; a summary of observations is given in Table  \ref{tab:obs}. 

All the observations were taken using the $300V$ grism, providing a spectral resolution of 12~\r{A} (as determined from arc lamp calibration frames). The spectropolarimetric data were reduced and analysed using our package For Use with Spectropolarimetry (FUSS\footnote[2]{Code available at: https://github.com/HeloiseS/FUSS}). FUSS is a combination of IRAF.cl and python scripts. 
The IRAF\footnote[3]{IRAF is distributed by the National Optical Astronomy Observatory, which is operated by the Association of Universities for Research in Astronomy (AURA) under a cooperative agreement with the National Science Foundation.} routines partially automate the extraction of spectra following the prescription of \cite{maund05bf}. 
The python modules of FUSS then calculate the Stokes parameters (for linear spectropolarimetry see \citealt{patat06}, for circular spectropolarimetry see \citealt{fors2}). The data are corrected for chromatic zero angles and polarisation bias \citep{wang97}. The Doppler correction and interstellar polarisation correction (see Section \ref{sec:isp}) were also performed with FUSS. In order to increase the level of signal-to-noise the spectropolarimetric data were re-binned to 45 \r{A}, which did not affect the resolution considering the breadth of the spectral features. The spectroscopic data obtained on 09 July 2014 were reduced with our IRAF.cl scripts following the standard procedure.

\section{RESULTS}
\label{sec:res} 
\subsection{Lightcurve}
\begin{figure}
\centering
\includegraphics[width=9cm]{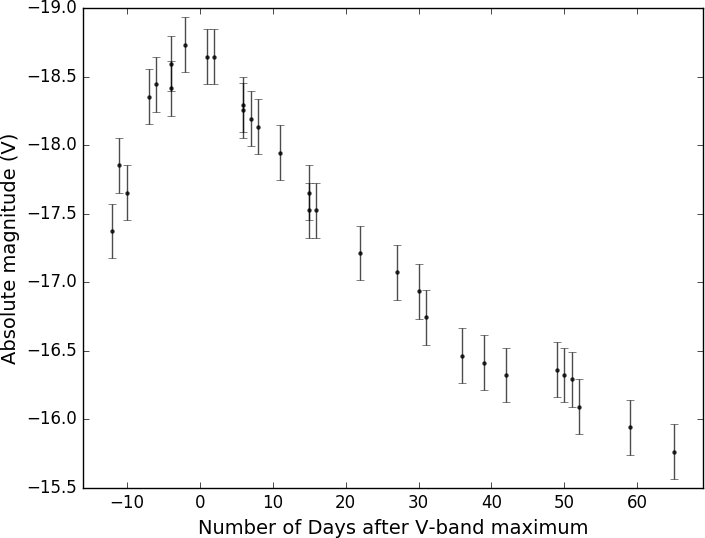}
\caption{\label{fig:LC} De-reddened lightcurve of SN~2014ad from 12 March 2014 to 28 May 2014. Details of photometric acquisition can be found in \cite{howerton17}.}
\end{figure}

V-band observations were acquired between 12 March 2014 and 28 May 2014 (for details see \citealt{howerton17}). A luminosity distance of 24.7 Mpc was calculated from the redshift of MRK1309 (z=0.005723, \citealt{HIPASS06}), and we estimated that SN~2014ad reached maximum with an absolute V-band magnitude of -18.8 $\pm$ 0.2 mag on 24 March 2014. All epochs subsequently mentioned in this paper are quoted with respect to this date. The absolute V-band magnitude was corrected for extinction using the Milky Way and host galaxy reddening we calculated (see section \ref{sec:isp}). An exhaustive analysis of the lightcurve is not the main focus of this paper, but we note that the rise time to maximum and the breadth of the peaks are similar to those observed in SNe 1998bw and 2003dh \citep{deng05} with an estimated $\Delta m_{15}$ \about 0.9 mag, which could suggest a similar ejecta mass (although see \citealt{wheeler15}).

\subsection{Flux Spectroscopy}
\label{sec:syn++}
\begin{figure}
\centering
\includegraphics[width=9cm]{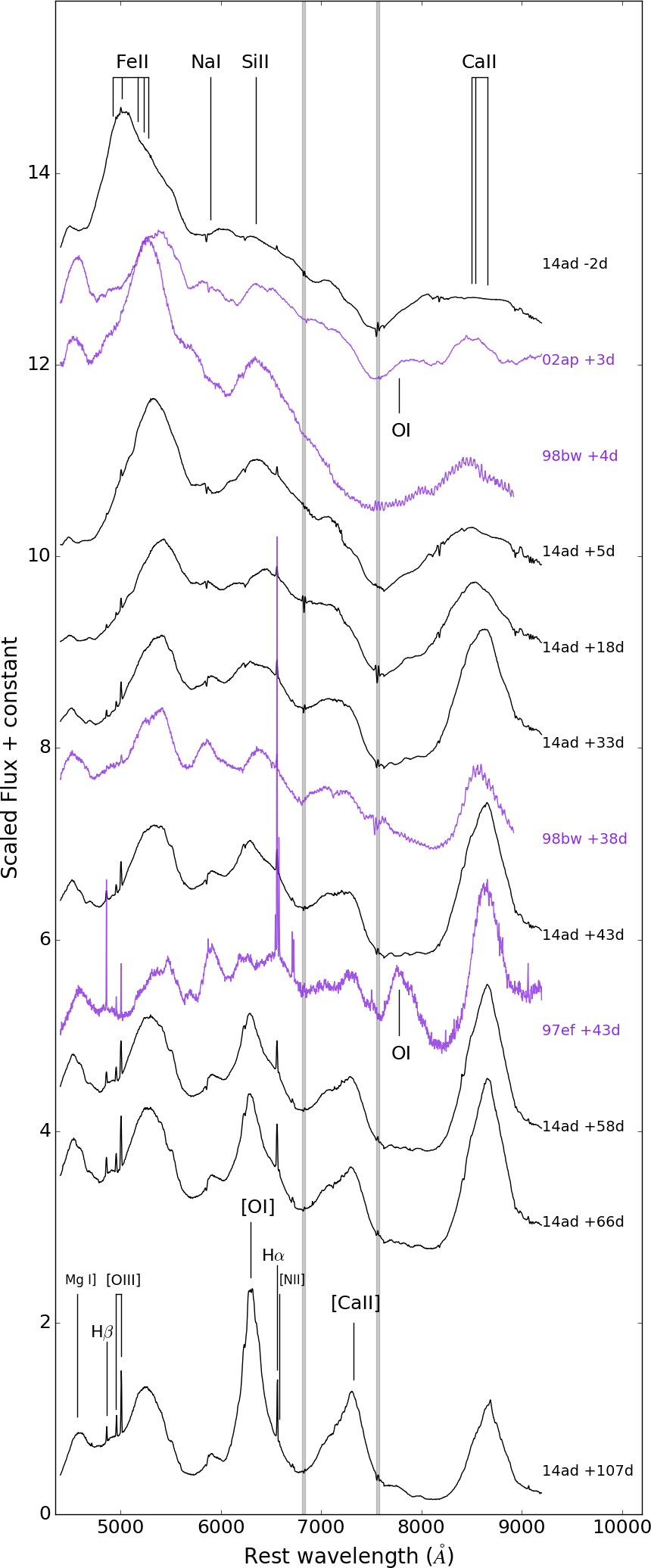}
\caption[ ]{\label{fig:spctr}Flux spectrum of SN~2014ad at 8 epochs (black) corrected for recessional velocity. We compare with the flux spectra of the three SNe Ic-bl which most closely matched SN~2014ad according to Gelato\footnotemark[4]: SN~1997ef (43 days after V-band maximum), SN~1998bw (at +4 days and +38 days), as well as SN~2002ap (3 days post V-band maximum).  The grey bands highlight telluric features. SN~1998bw is known for its connection to GRB 980425 \citep{woosley99}, SN~1997ef was thought to be the optical counterpart of GRB 971115, but the significance of their correlation is much lower than for SN~1998bw/GRB 980425 \citep{iwamoto00}. The narrow lines in the spectra of SN~2014ad are associated with the interstellar medium of the host galaxy.}
\end{figure}
\footnotetext[4]{https://gelato.tng.iac.es/}

The spectral features of SN~2014ad are very broad, particularly at early times, which is characteristic of SNe Ic-bl (see Figure \ref{fig:spctr}). At 5 days after V-band maximum, SN~2014ad shows wider features than SN~2002ap at +3 days, that are nearly as broad as SN~1998bw at +4 days. Although the spectral lines become less broad as time passes, SN~2014ad still shows features that are much wider than that of SN~1997ef 43 days after V-band maximum. Additionally, at +33d and +43d SN~2014ad is very similar to SN~1998bw at +38d.

\begin{figure}
\centering
\includegraphics[width=8.5cm]{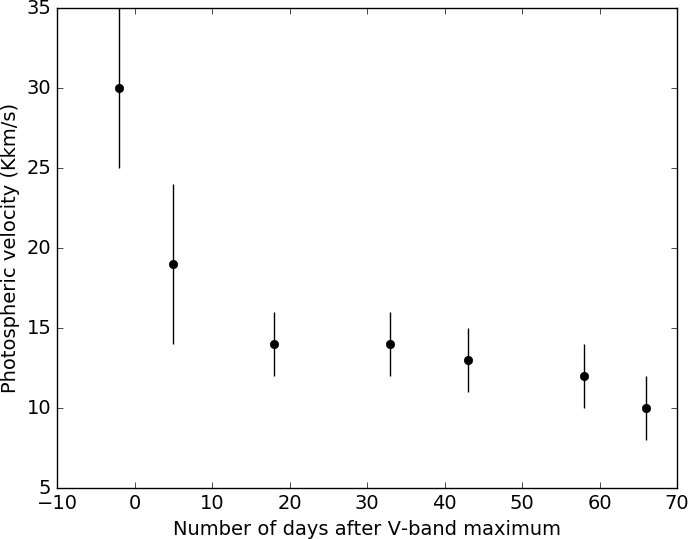}
\caption{\label{fig:phot_vel} Time evolution of the photospheric velocity of SN~2014ad, as measured by fitting the spectra using SYN++}
\end{figure}

In CCSNe, the line centre velocity of the Fe\,{\sc ii} $\lambda$5169 absorption component is commonly used as proxy for the photospheric velocity (e.g. \citealt{modjaz16}). The extreme line blending occurring in the spectra of SN~2014ad, however, made using this feature unreliable \citep{liu16}. Consequently, we used \textsc{syn++} \citep{syn++} to fit the Fe\,{\sc ii} blend in our spectra of SN~2014ad and determine the photospheric velocity, as well as confirm line identification. \textsc{syn++}  is a radiative transfer code that assumes spherical symmetry and no electron scattering. It allows the user to choose the ions to be added to the synthetic spectra, and change parameters such as the photospheric velocity, the opacity, the temperature and the velocity of each ion, in order to construct the best fit to the data. We found a photospheric velocity as high as 30,000 $\pm$ 5,000~\kms at -2 days, decreasing to 10,000 $\pm$ 2,000~\kms by +66 days (see Figure \ref{fig:phot_vel}). 

\begin{figure}
\centering
\includegraphics[width=8.5cm]{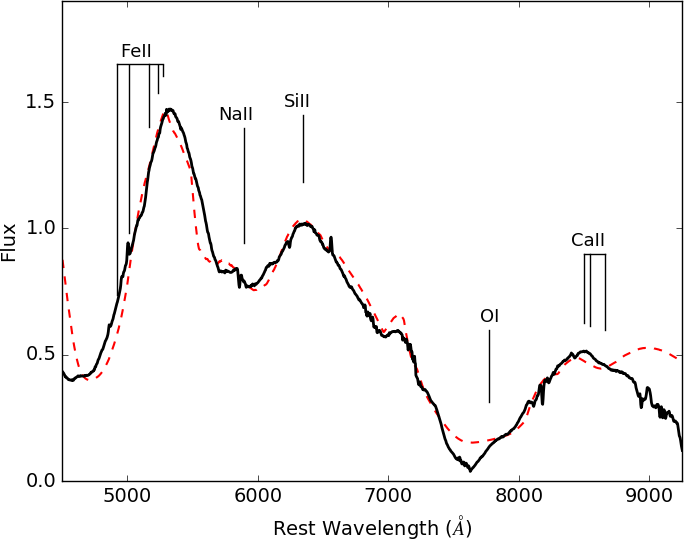}
\caption{\label{fig:syn++} Spectrum of SN~2014ad at 5 days after V-band maximum (solid black) and model obtained using \textsc{syn++} (dashed red). The model includes Fe\,{\sc ii}, Na\,{\sc i}, Si\,{\sc ii}, O\,{\sc i} and Ca\,{\sc ii}.}
\end{figure}

The extreme line blending caused by the high ejecta velocities also made line identification challenging. At all epochs, best \textsc{syn++}  fits were obtained for synthetic spectra containing Fe\,{\sc ii}, Na\,{\sc i}, Si\,{\sc ii}, O\,{\sc i} and Ca\,{\sc ii}. It was not possible to obtain uniformly good fits, especially at later epochs due to the appearance of forbidden lines of Mg\,{\sc i}], [O\,{\sc i}] and [Ca\,{\sc ii}]. We were however able to find good fits of the Fe\,{\sc ii} complex at all epochs; Figure \ref{fig:syn++} shows our fit to the spectrum of SN~2014ad at +5 days. 
 
At the earlier epochs, the spectrum is dominated by a blend of iron lines in the region 4500-5500~\ang, and a deep absorption component with a minimum at \about 7500~\ang\, corresponding to a blend of O\,{\sc i} $\lambda$7774 and Ca\,{\sc ii} infrared (IR) triplet. In the first epoch the emission component of the Ca\,{\sc ii}\,+\,O\,{\sc i} blend is flat topped, but it becomes prominent by +5 days. At this epoch, a strong Si\,{\sc ii} $\lambda$6355 emission has also emerged. Over time, line blending diminishes as opacity decreases and we are probing the deeper, slower ejecta, however the O\,{\sc i} $\lambda$7774 never completely separates from the Ca II IR triplet as it does in SN~1997ef. At later epochs, the strength of the iron blend progressively decreases, and the Ca\,{\sc ii} IR triplet emission strengthens. By 43 days after maximum, we can see the spectrum of SN~2014ad starting to transition from the photospheric phase towards the nebular phase: The [O\,{\sc i}] emission at \about 6300~\ang\,  dominates over the Si\,{\sc ii} $\lambda$6355 emission observed at previous epochs, and the absorption of the  Ca\,{\sc ii} IR triplet becomes increasingly flat-bottomed as the P-Cygni profile fades and is replaced by just an emission feature. At +107 days, SN~2014ad has a typical nebular spectrum, with (semi-)forbidden lines of Mg\,{\sc I}] $\lambda$4571, [O\,{\sc I}] $\lambda$6300 and [Ca\,{\sc II}] $\lambda\lambda$7291,7824.

\subsection{Linear Spectropolarimetry}
\label{sec:lin.specpol}

%

\begin{figure}
\centering
\includegraphics[width=8.5cm]{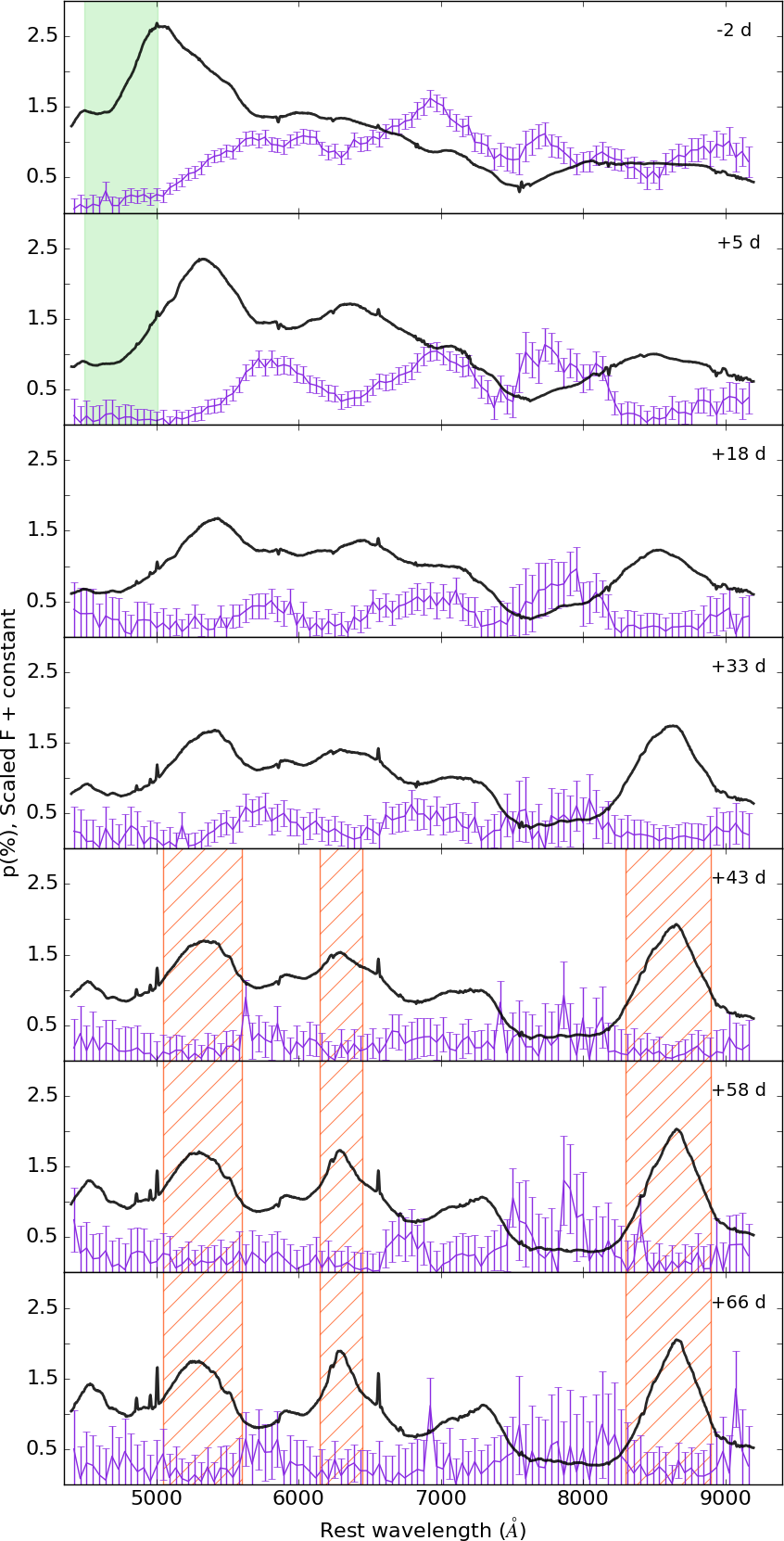}
\caption{\label{fig:flu_n_pol} Flux spectra (black) and the degree of polarisation (purple) of SN~2014ad from $-2$ days to 66 days after V-band maximum. The polarisation spectra were binned to 45~\ang\, in order to increase the signal to noise ratio. The data presented here are not corrected for ISP. The ranges used to determine the ISP are indicated by the shaded green region -- method (i) -- and hashed orange region-- method (ii). For more detail see Section \ref{sec:isp}. For line IDs see Figure \ref{fig:spctr}.}
\end{figure}

The degree of polarisation $p$ and Stokes $I$ for each of our 7 epochs of linear spectropolarimetry are shown in Figure \ref{fig:flu_n_pol}. At \mbox{-2} days, 5 peaks can be seen in the polarisation spectra at \about5760~\ang\, ($p=1.07 \pm 0.10 \%$), 6030~\ang\, ($p=1.09 \pm 0.10 \%$), 6925~\ang\, ($p= 1.62 \pm 0.12 \%$),  7730~\ang\, ($p= 1.12 \pm 0.16$), and 8130~\ang\, ($p= 0.86 \pm 0.14 \%$). Note that the values of polarisation in brackets are the values recorded for the wavelength bin at each peak. The 6925~\ang\, peak shows the highest level of polarisation recorded in our data set. This feature is likely the result of O\,{\sc i} $\lambda$~7774 at -32 700~\kms. As for the other features, the 5760~\ang\, peak in polarisation is associated with Si\,{\sc ii} $\lambda$6355 with velocity -28 000~\kms, the 7730~\ang\, feature arises from Ca\,{\sc ii} at -29 300~\kms, and the peak seen at 8130~\ang\, could be O\,{\sc i} $\lambda$9264 at -36 600~\kms. The origin of the 6030~\ang\, feature, however, remains unclear. The photospheric velocity at -2 days found using \textsc{syn++} was 30,000 $\pm$ 5,000~\kms (see Section \ref{fig:phot_vel}), therefore the line-forming region yielding the Ca\,{\sc ii} IR, Si\,{\sc ii} and O\,{\sc ii} $\lambda$7774 polarisation features must be close to the photosphere at the first epoch.

By 5 days post-maximum, the 5760 and 6030~\ang\, features have merged, yielding a broader peak centred on 5770~\ang\, with polarisation $p=0.86\pm0.01 \%$ (calculated as the average polarisation between  5670 and 5970 \ang). Similarly, the 7730~\ang\, (Ca\,{\sc ii}) and 8130~\ang\, (O\,{\sc i}) peaks seen in the first epoch are replaced by a much broader feature extending between \about 7580~\ang\, and \about 8130~\ang, with $p=0.86\pm0.15 \%$ (the average between 7580 and 8130\ang). The O\,{\sc i} $\lambda$7774 peak is still present, but has moved \about50~\ang\, to the red (now at 6970~\ang) and its polarisation has decreased to $p=1.04\pm0.13 \%$. The first peak at \about 5800~\ang\, is consistent with Si\,{\sc ii} $\lambda$6355 at -26 000~\kms, and the O\,{\sc i} $\lambda$7774 velocity is now -31 000 \kms, therefore both elements show a decrease in velocity of \about 2000~\kms. The photospheric  velocity at this epoch was found to be 19000 $\pm$ 5000~\kms (see Figure \ref{fig:phot_vel}), suggesting that the line-forming region is more detached from the photosphere by the second epoch than in the first epoch.

The amplitude of the Si\,{\sc ii} $\lambda$6355 and O\,{\sc i} $\lambda$7774 polarisation peaks decreases drastically by 18 days after V-band maximum, with $p=0.51\pm0.2 \%$ and $p=0.65\pm0.2 \%$ respectively. The Ca\,{\sc ii} IR/O\,{\sc i}~$\lambda9264$ peak, however, remains constant with $p=0.96\pm0.3 \%$. Its degree of polarisation then decreases to \about0.5\% by +33 days, and is dominated by noise, whereas the first 2 peaks remain constant. By 43 days after V-band maximum, and at later epochs, the polarisation spectra are dominated by noise, as the supernova has faded considerably.

\subsection{Circular Spectropolarimetry}
\label{sec:circ_pol}

\cite{wolstencroft72} suggested that if the core of a progenitor collapses to a neutron star it may have sufficiently large magnetic fields to induce detectable circular polarisation from bremsstrahlung radiation or synchrotron emission. They cautioned that the degree of circular polarisation would be significantly influenced by the precise conditions of the explosion, and no strong evidence supporting the presence of circular polarisation in CCSNe has been detected so far. 

\begin{figure}
\centering
\includegraphics[width=8.5cm]{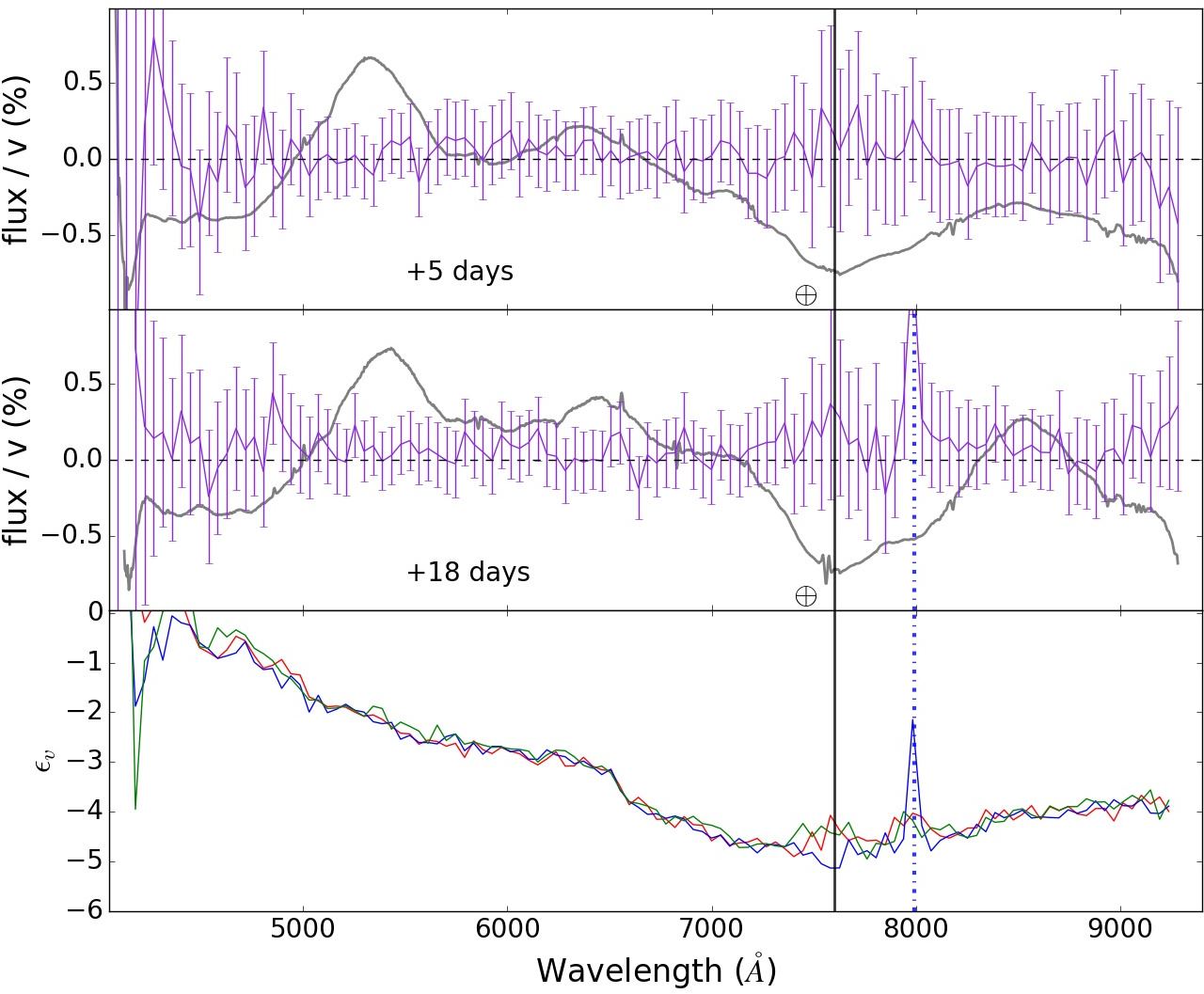}
\caption{\label{fig:circ_pol}Circular polarisation binned to 45~\ang of SN~2014ad on 2014 March 29 (+5 days) and 2014 April 11 (+18 days) and the instrumental signature correction $\epsilon_v$. The spectrum at the corresponding epochs are under-plotted in grey. On the plot of $\epsilon_v$ the red line corresponds to +5 days and the blue and red line  are associated with the two data sets obtained at +18 days. The black solid vertical line indicates the location of the telluric line occurring at 7600-7630~\ang, and the blue dashed line at 7996~\ang\, shows the correlation of the discrepancies observed in $\epsilon_{v}$ and v at +18 days (see Section \ref{sec:circ_pol}).}
\end{figure}

We investigated the potential presence of circular polarisation in SN~2014ad. Circular spectropolarimetry was acquired on 29 March 2014 and 11 April 2014. At both epochs we found that the average circular polarisation (0.023 and 0.068 at epoch 2 and 3, respectively) was lower than the standard deviation (0.116 and 0.115 at epoch 2 and 3, respectively) over the range 4500-9300~\ang. The slight variability around \about7600~\ang\,  is attributed to the strong telluric line occurring at these wavelengths and the outlier at \about 8000~\ang\, seen at +18 days correlates with a spurious spike in the  instrumental signature correction $\epsilon_v$ \citep{maund08}. We therefore concluded that the data are consistent with a null circular polarisation (see Figure \ref{fig:circ_pol}). 

\section{ANALYSIS}
\label{sec:analysis}

\subsection{Spectropolarimetry}


\subsubsection{Interstellar Polarisation}
\label{sec:isp}
The dust present in the interstellar medium along the line-of-sight may induce polarisation. The resulting interstellar polarisation (ISP) must be quantified in order to isolate the polarisation that is intrinsic to the SN. An upper limit on the ISP can be constrained if we assume a standard Serkowski-Galactic type ISP \citep{serkowski73}. The degree of polarisation due to the ISP ($p_{\text{ISP}}$) is then related to the total reddening $E(B-V)_{\text{total}}$ = $E(B-V)_{\text{MW}}$ + $E(B-V)_{\text{host}} $  by the relation: 
\begin{equation}
p_{\text{ISP}} (\%) \le 9 \times E(B-V)_{\text{total}}.
\end{equation}
The reddening can be estimated from the Na\,{\sc i} D lines in the spectra following \citet{Poznanski12}. In our spectra of SN~2014ad, only the Na\,{\sc i} D line arising from the Milky Way contribution is visible. In order to put an upper limit on the reddening occurring in MRK 1309, we calculated a detection limit at  3$\sigma$ assuming a spectral resolution of FWHM = 12~\ang\, for Na\,{\sc i} D at the recessional velocity of the host galaxy. Our estimates for the reddening are $E(B-V)_{\text{MW}}$ = 0.14 mag, and $E(B-V)_{\text{Host}} \le$ 0.023 mag, yielding an upper limit for the ISP $p_{\text{ISP}} \le$ 1.56 \%. By applying the Serkowski law to the host galaxy we implicitly assumed that the size and composition  of the dust in the host galaxy are the same as that of the Milky Way, which may not be true.

The Milky Way component of the ISP can also be evaluated by finding Milky Way stars near the line-of-sight towards SN~2014ad. If we assume that their intrinsic polarisation is null, then the polarisation measured for these stars is purely due to the Galactic ISP. Within 2 degrees of SN~2014ad, two stars were found in the \cite{Heiles} catalogue, with $p =$ 0.08 ($\pm$ 0.035) \% and 0.11 ($\pm$ 0.066)~\%. The first star (HD 104304) has a known distance of 12.76 parsec \citep{hd104304}, and it therefore does not sample the full Galactic dust column. The other star (HD 104382) does not have a known parallax. With a Galactic latitude of 50.5\degree, however, SN~2014ad is located far from the Galactic plane, and we do not expect significant dust alignment along the line-of-sight. Consequently we concluded that a lower limit for the Milky Way component of the ISP is $\sim$ 0.1 \%. 

In order to directly quantify the ISP from the polarisation data obtained for SN~2014ad, assumptions must be made. We use two different assumptions to calculate two independent estimates of the ISP: complete depolarisation in line blanketed regions and complete depolarisation at late times.
\begin{enumerate}
\item[(i)]Strong depolarisation is observed at short wavelengths (see Figure \ref{fig:flu_n_pol}) due to line blanketing by Fe\,{\sc ii} and Sc\,{\sc ii} lines \citep{wang01}, and if we assume complete depolarisation of the SN~light, the resulting observed degree of polarisation is $p_{\text{ISP}}$. We calculated the polarisation at -2 days and +5 days over the wavelength range 4500-5000 \r{A} (the green shaded regions in Figure \ref{fig:flu_n_pol}), yielding $q_{\text{ISP}}$ = 0.20 ($\pm$0.07)\%, $u_{\text{ISP}}$ = 0.095 ($\pm$0.09)\%  and $p_{\text{ISP}}$ = 0.22 ($\pm$0.07)\% at -2 days and $q_{\text{ISP}}$ = 0.079 ($\pm$0.09)\%, $u_{\text{ISP}}$ = 0.013 ($\pm$0.05)\% and $p_{\text{ISP}}$ = 0.08 ($\pm$0.08)\% at +5 days. Because ISP should be constant with time, we average the two sets of Stokes parameters: $q_{\text{ISP}}$ = 0.14 ($\pm$0.06)\% and $u_{\text{ISP}}$ = 0.04 ($\pm$0.05)\%, which correspond to $p_{\text{ISP}}$ = 0.15 ($\pm$0.08)\% and a polarisation angle $\theta_{\text{ISP}}$ = 10\degree ($\pm$30\degree), which is within the limits previously established. We considered the line blanketing regions at the first two epochs only, because the higher ejecta velocities result in greater line blending and their spectropolarimetric data have better levels of signal-to-noise as SN~2014ad was at its brightest.\\

\item[(ii)]At late times, when the ejecta transitions to the optically thin nebular phase (in the case of SN 2014ad, from +43 days), electron scattering does not dominate anymore and the intrinsic polarisation of the supernovae is expected to drop to zero. Consequently, if we assume complete depolarisation in the SN ejecta at late times, any level of polarisation observed must be due to ISP. Because SN 2014ad had significantly faded at +43, +58 and +66 days, the levels of signal to noise are lower than in early epochs. We focus our analysis on the polarisation that correlates with strong emission lines as the higher levels of flux result in smaller errors in the polarisation, and because emission lines are often associated with depolarisation even when electron scattering is significant, although see \cite{kasen03}. With this in mind, we pick 3 wavelength ranges (5050-5600~\r{A}, 6150-6450~\r{A} and 8300-8900~\r{A};  see hashed orange region in Figure \ref{fig:flu_n_pol}) corresponding to the \mbox{(semi-)}forbidden lines of magnesium, oxygen and calcium, and we average the Stokes parameters across these ranges. Doing this for epochs 5, 6 and 7 yielded 9 values for $q_{\text{ISP}}$ and $u_{\text{ISP}}$, which were averaged, yielding $q_{\text{ISP}}$ = 0.05 ($\pm$0.05)\%, $u_{\text{ISP}}$ = 0.15 ($\pm$0.12)\%, $p_{\text{ISP}}$ = 0.16 ($\pm$0.13)\% and $\theta_{\text{ISP}}$ = 36\degree ($\pm$5\degree). These values are consistent with the ones found using the line blanketing assumption.
\end{enumerate}

The two methods described above therefore resulted in ISP estimates that were small and consistent with each other. We decided to adopt the Stokes parameters values calculated with the first method for our subsequent ISP corrections since they were estimated using polarisation spectra at early days with better signal to noise ratio.


\subsubsection{Polarisation in the Stokes q-u plane}
\label{sec:q-u}
\begin{landscape}
\begin{figure}
\centering
\includegraphics[width=24cm]{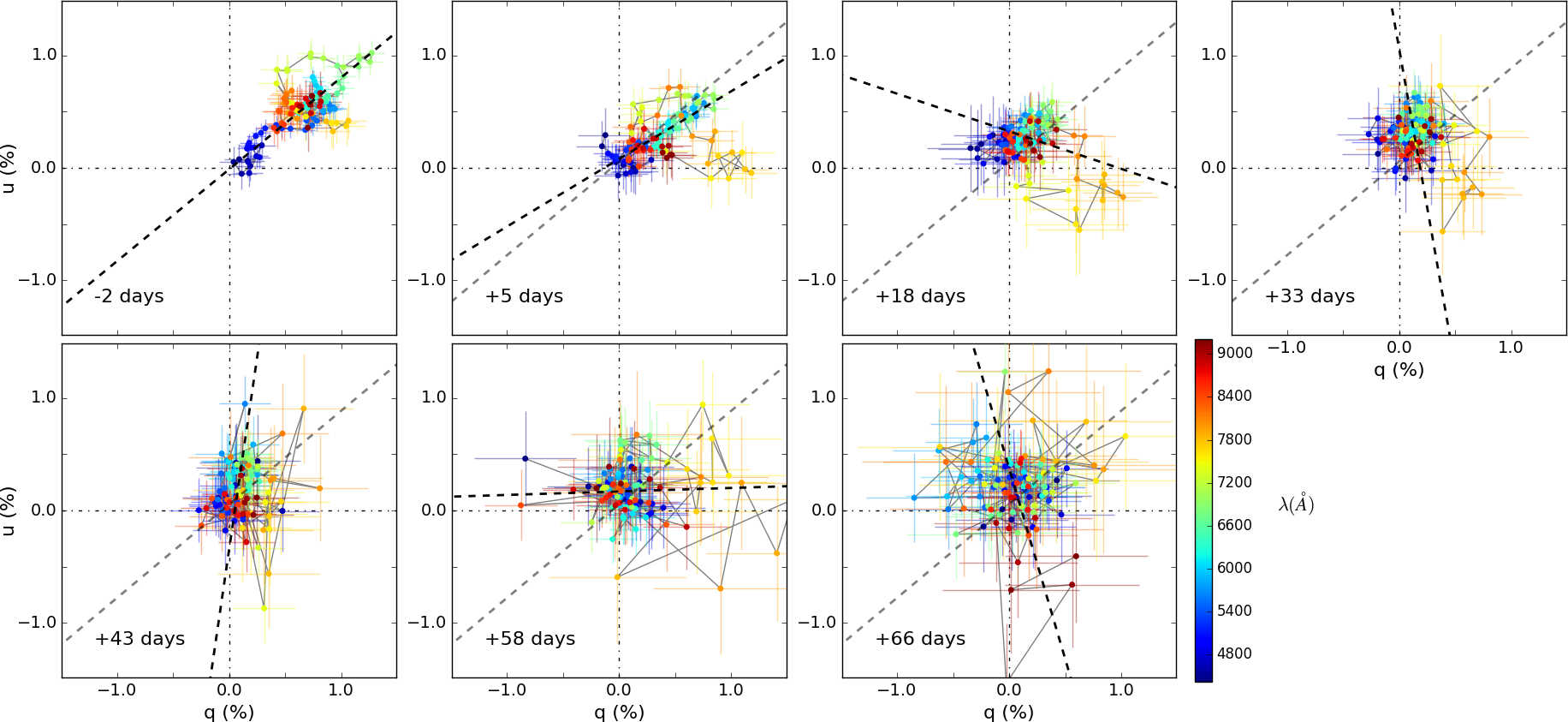}
\caption{\label{fig:qu}Stokes $q-u$ planes of SN~2014ad in 7 epochs. The polarisation data were binned to 45 \ang and the colour scheme represents wavelength. The dominant axis was calculated at each epoch using the Orthogonal Distance Regression (ODR) method, and is shown by the overlaid black dashed line. The grey dashed line is the dominant axis found at -2 days superposed on the data of subsequent epochs for comparison purposes. The data presented here are corrected for ISP using the values derived in Section \ref{sec:isp}.}
\end{figure}
\end{landscape}

The ISP corrected data were plotted on the Stokes $q-u$ plane for each epoch as shown in Figure \ref{fig:qu}, where the colour scale represents wavelength. 

At 2 days before V-band maximum the data are aligned along a very clear dominant axis with P.A. $= 21.5^{\circ} \pm 0.5 ^{\circ}$ (corresponding to an angle of 43\degree on the $q-u$ plane) as shown by the over-plotted Orthogonal Distance Regression (ODR\footnote[5]{\textsc{python} package: https://docs.scipy.org/doc/scipy/reference/odr.html}) fit (see Figure \ref{fig:qu}). All ODR fits were performed using the entire wavelength range. A loop at \about7000~\ang\, can be seen following the dominant axis orientation. From our identification in Section \ref{sec:lin.specpol}, this feature is associated with O\,{\sc i} $\lambda$7774. 

Our data at +5 days show that the P.A. of the dominant axis (P.A $= 19.5^{\circ} \pm 1.0^{\circ}$) is similar to the first epoch, and both dominant axes are consistent with the data. The loop at \about 7000~\ang\, has become less prominent, but is still distinguishable. Another loop arises at \about 7700~\ang\, corresponding to the Ca\,{\sc ii} IR triplet, which runs almost perpendicular to the dominant axis. This indicates a significant departure from axial symmetry in the distribution of calcium in the form of large clumps.

By +18 days the presence of a dominant axis is less pronounced as the overall degree of polarisation decreases, but a Pearson test performed on the data -- excluding the Ca II loop -- yielded a coefficient of 0.52, supporting the presence of a linear correlation. The ODR fit included the Ca II IR component and seems to follow the direction of the strong calcium loop, but it is not an accurate representation of the data as a whole. With the exception of the Ca\,{\sc ii} IR triplet, the data follow a direction similar to that of the dominant axis found in the first epoch. The O\,{\sc i} $\lambda$7774 loop has completely disappeared, but the calcium feature is still very strong. In later epochs, the calcium loop weakens and the data as a whole cluster towards the origin as the overall level of polarisation decreases and the noise dominates. The ODR fits of epoch 4 to 7 were plotted on their respective $q-u$ plane for completeness, but the Pearson correlation calculated for the data at these dates (\mbox{-0.08}, 0.13, 0.04 and -0.03, respectively) revealed that the presence of a linear correlation is very unlikely.

\subsubsection{Rotation of the Stokes parameters}
\label{sec:rot_stokes}
\begin{figure*}
\centering
\includegraphics[width=17cm]{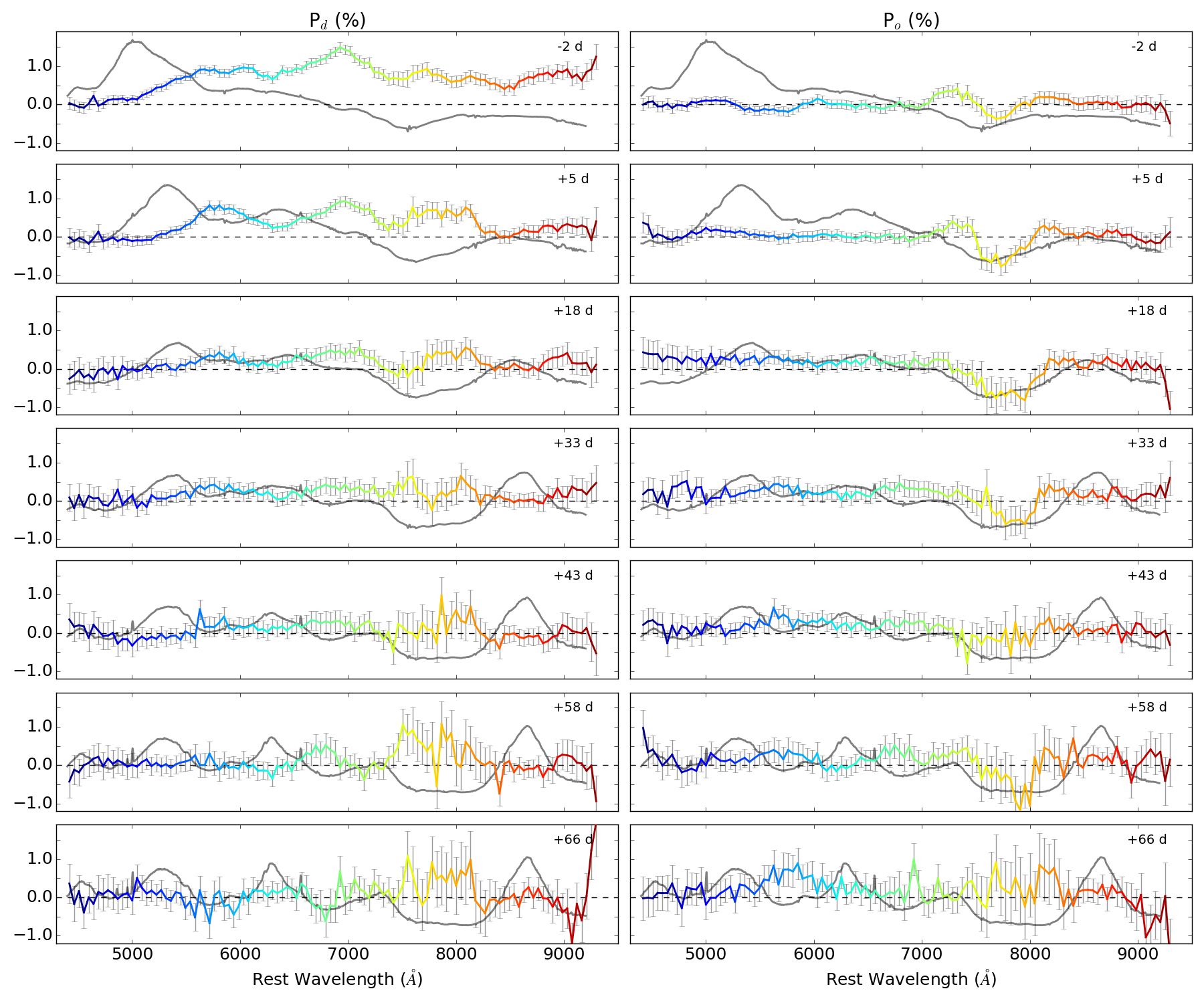}
\caption{$P_d$ and $P_o$ (multicolor), as defined by Equation \ref{equ:rot}, corresponding to our 7 epochs of spectropolarimetry. The spectropolarimetric data were binned to 45 \ang~and corrected for the ISP before being rotated. The flux spectrum (unbinned) is plotted (grey) at each epoch. For line IDs see Figure \ref{fig:spctr}. }
\label{fig:Pd_Po}
\end{figure*}

The data presented on the $q-u$ plane can be rotated  in order to obtain two new components which correspond to the Stokes parameters projected onto the dominant axis ($P_d$) and the axis orthogonal to the dominant axis ($P_o$). If we consider the rotation matrix of the form:
\begin{equation}
\mathbf{R(\theta_{\text{rot}})}=
\begin{pmatrix}
\cos\theta_{\text{rot}} & -\sin\theta_{\text{rot}} \\
\sin \theta_{\text{rot}} & \cos\theta_{\text{rot}}\\
\end{pmatrix}
\end{equation}
then the rotated Stokes parameters are given by:
\begin{equation}\label{equ:rot}
\begin{pmatrix}
P_d \\
P_o\\
\end{pmatrix}
=\mathbf{R(\theta_{\text{rot}})} \cdot
\begin{pmatrix}
q\\
u\\
\end{pmatrix}
\end{equation}
where $\theta_{\text{rot}} = -2 \times \theta_{\text{dom}}$. 

The direction of the dominant axis found using ODR at the first epoch is consistent with the data at epochs 2 and 3, and is therefore chosen to perform the rotation, corresponding to \mbox{$\theta_{\text{rot}}$ = 43\degree}. The resulting rotated Stokes parameters $P_d$ and $P_o$ were then plotted against wavelength, as shown in Figure \ref{fig:Pd_Po} (the colour scale matches that used in Figure \ref{fig:qu}).

Overall, at each epoch the orthogonal parameter $P_o$ is within 1$\sigma$ of zero along most of the spectrum, except within the wavelength ranges corresponding to loops seen in the $q-u$ plane (see Figure \ref{fig:qu}). This shows that our choice of dominant axis is appropriate and confirms our interpretation that the ODR fit of the data at +18 days was dominated by a prominent loop rather than the direction of the dominant axis consistent with the majority of the data. In Figure \ref{fig:qu}, one can see that the data at +18 days are sightly offset with respect to the chosen dominant axis (indicated grey dotted line -- identical to the dominant axis found using ODR in the first epoch) but seems parallel. This behaviour is also seen in Figure \ref{fig:Pd_Po} at +18 days, where in $P_o$ most of the data (apart from the Ca\,{\sc ii} loop between \about 7200-8200~\ang) run parallel to the dominant axis ($P_o$=0\%) but is shifted up by about 0.2\%.

The significant departures from null polarisation in $P_o$ correspond to loops in the $q-u$ plane. At -2 days, the main feature is found between \about 7000~\ang\, and \about 8300~\ang. It deviates from the dominant axis by \about 0.5\% on either side, and is most likely caused by a mixture of O\,{\sc i}~$\lambda$7774 and Ca\,{\sc ii}~$\lambda$8567. At +5 days the deviation from the dominant axis associated with O\,{\sc i}+Ca\,{\sc ii} becomes more prominent between 7500~\ang\, and 8000~\ang, but still extends from \about 7000~\ang\, and up to \about 8300~\ang. By +18 days the loop only departs on one side of the dominant axis and now starts at \about 7200~\ang, which may indicate that the oxygen component has reduced significantly and the feature is now dominated by the Ca\,{\sc ii} component. Two weeks later, the amplitude of the loop has started to decrease and by 43 days after V-band maximum it is not distinguishable anymore.

\subsubsection{Loops in the $q-u$ plane}
\label{sec:loops}
\begin{figure*}
\centering
\includegraphics[width=17cm]{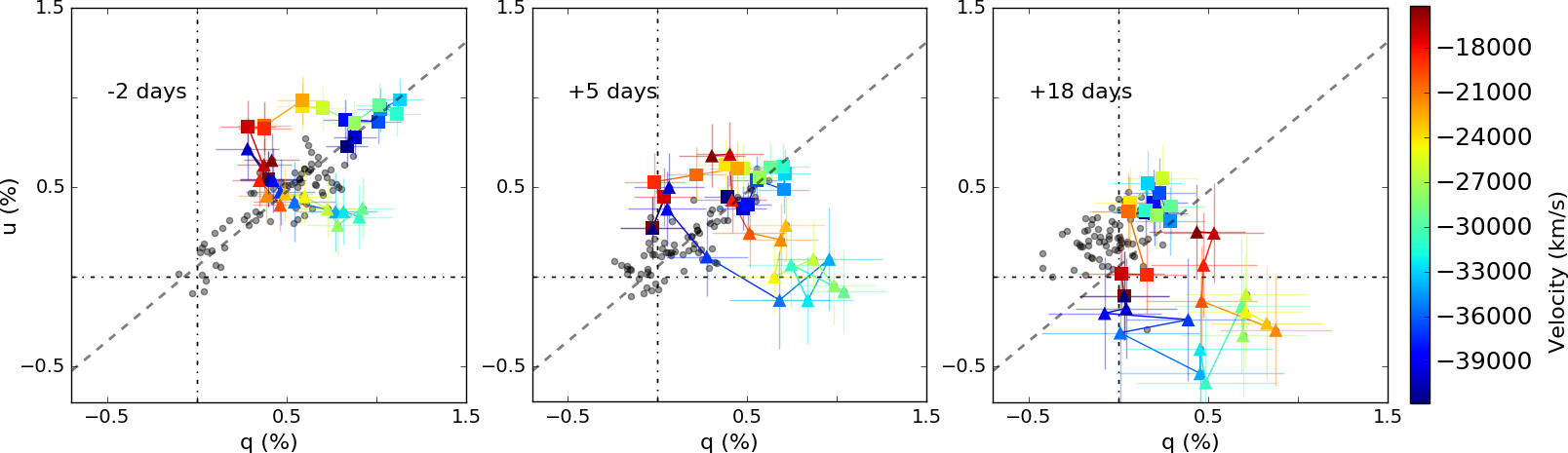}
\caption{Stokes $q-u$ planes of SN~2014ad in the first three epochs (see also Figure \ref{fig:qu}); the spectropolarimetric data were binned to 45 \ang. The colour scheme represents ejecta velocity. The dominant axis (dashed grey line) is the dominant axis as calculated by ODR at the first epoch. The data marked by rectangles correspond to O\,{\sc i} $\lambda$7774 and the data marked by triangles is Ca\,{\sc ii} IR. For comparison the data over the range 4500-9300\ang\, were plotted (grey points). For clarity error bars were only plotted for the O\,{\sc i} $\lambda$7774 and the Ca\,{\sc ii} IR data. The data were corrected for ISP (see Section \ref{sec:isp}). }
\label{fig:loops}
\end{figure*}

In Figure \ref{fig:loops} we show the O\,{\sc i} $\lambda$7774 and Ca\,{\sc ii} IR loops on the $q-u$ planes at -2, +5 and +18 days, where the colour scale now represents ejecta velocity. 
At -2 days the O\,{\sc i} feature dominates and the Ca\,{\sc ii} IR feature forms a line oriented away from the dominant axis. By +5 days the Ca\,{\sc ii} IR feature strengthens and becomes a loop oriented in the same direction as the line observed at -2 days, while the oxygen loop becomes less prominent. At +18 days the O\,{\sc i} data is in the same locus as the majority of the data, and the Ca\,{\sc ii} IR loop has strengthened again. At -2, +5 and +18 days, the oxygen and calcium loops are consistently formed anti-clockwise with increasing wavelength on the q-u plane. 

Furthermore, the oxygen and calcium features never coincide simultaneously in their Stokes parameters and in velocity space, which would indicate overlap of the line-forming regions. This suggests that the line-forming regions of O\,{\sc i} and Ca\,{\sc ii} are very distinct from each other both in radial velocity (= radius) and on the plane of the sky.

\subsection{V-band Polarisation}
\label{sec:vpol}
Because the spectral features of SN 2014ad were so broad, we could not isolate a line-free region of the spectrum to estimate the continuum polarisation. We therefore followed \cite{leonard06} who use use V-band polarimetry as a proxy for the overall evolution of the polarisation. The V-band polarisation of SN~2014ad was calculated from the spectropolarimetric data by weighting all our spectra by the transmission function of the Johnson V filter before performing the polarimetry calculations. \textsc{pysynphot}\footnote[6]{https://pysynphot.readthedocs.io/en/latest/} \citep{pysynphot} was used to calculate the weighted spectrum for each ordinary and extra-ordinary ray, which were subsequently processed with FUSS to calculate the resulting degree of polarisation and normalised Stokes parameters.

\begin{figure*}
\centering
\includegraphics[width=17cm]{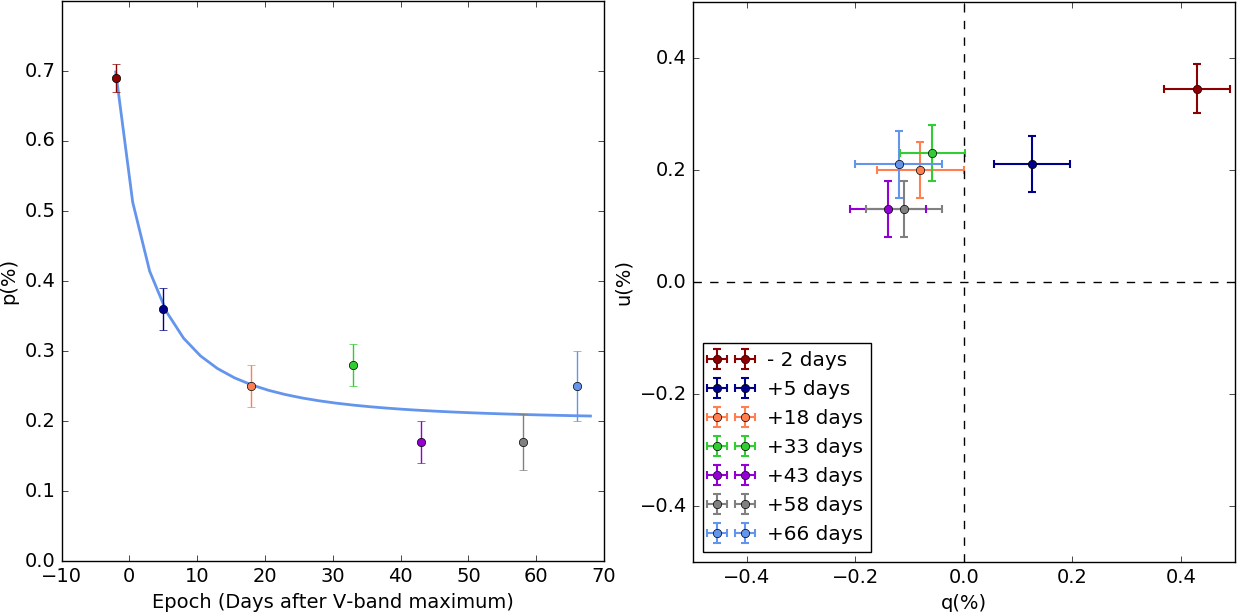}
\caption{\label{fig:pv} \textbf{Left panel:} Degree of V-band polarisation against time (i.e. epoch in days after V-band maximum). The blue line is a fit of the form $p$ = 40.3 t$^{-2}$ + 0.25, yielding a reduced $\chi^2$ = 7. The colours represent different epochs (see legend on right panel). \textbf{Right panel:} V-band polarisation of SN~2014ad in the $q-u$ plane. The colours represent different epochs (see legend).}
\end{figure*}

The resulting V-band polarisation for all 7 epochs is plotted against time and on the $q-u$ plane, see Figure \ref{fig:pv}. Were the decrease in the degree of polarisation caused solely by dilution due to the expansion of the ejecta, one would expect the polarisation to follow $p = \alpha \times t^{-2} + \beta$, where $t$ is time and $\alpha$ and $\beta$ are constants.

We tentatively fitted this relationship to our V-band polarisation using Monte Carlo methods to find the best values of $\alpha$ and $\beta$. For the best fit the reduced $\chi^2$ is 7, and the constants are found to be $\alpha$ = 40.3 $\pm$ 0.3 and $\beta$ = 0.250 $\pm$ 0.002. The fit is plotted as a blue line in Figure \ref{fig:pv}. The V-band polarisation of SN~2014ad seems to exhibit a behaviour that is not inversely proportional to time squared, which would indicate that the decrease in polarisation is not just a manifestation of ejecta expansion \citep{leonard06}. Given the size of the error bars, however, a t$^{-2}$ relation cannot be excluded.  

As seen in the right panel of Figure \ref{fig:pv} the polarisation angle remains nearly constant between the first two epochs (19 $\pm$ 3\degree and 29 $\pm$ 5\degree), then increases at epoch 3 (56 $\pm$ 4\degree), and subsequently stays constant. This behaviour could reflect a difference in geometry between the outermost layers, which are probed at early times (-2 and 5 days), and deeper layers of the ejecta, which are observed at later times (from +18 days through to +66 days) as the photosphere recedes through the envelope. This interpretation should however be considered with caution as the broad band-pass used here encompasses both continuum and line polarisation. 

\section{Discussion}
\label{sec:disc}
\subsection{Evolution of the shape of SN 2014ad}
\label{sec:shape}

If we assume as a first approximation that the photosphere is ellipsoidal, then continuum polarisation is related to the major to minor axis ratio of the ejecta and the line polarisation is caused by asymmetries in the distribution of the line-forming regions. Quantifying the level of continuum polarisation of SN~2014ad was made difficult by the extreme line blending, but with a maximum polarisation $p_{\text{max}}$ = 1.62 $\pm$0.12 \% the axis ratio of the photosphere could not be smaller than 0.72 $\pm$ 0.02 \citep{hoflich91}. A clear dominant axis is visible in the $q-u$ plots at -2 and +5 days, implying that the ejecta possess a strong axis of symmetry at early days. By +18 days, the data has receded towards the origin of the $q-u$ plots as the polarisation of SN~2014ad has decreased (see Figure \ref{fig:flu_n_pol}), but a dominant axis is still present (section 4.1.2 and 4.1.3). The overall decrease in $p$ and the disappearance of the dominant axis seem to indicate that the deeper ejecta are more spherically symmetric than the outer envelope, with the data at +33 days indicating a maximum axis ratio of 0.89 $\pm$ 0.01. The $q-u$ plots of the last 3 epochs do not offer more insight as the data   are dominated by noise. 

Significant departures from the identified dominant axis can be seen in the form of loops in the $q-u$ plots at epoch 1 through 4, indicating that the line-forming regions are made of large clumps that are asymmetrically distributed.  At -2 days features of O\,{\sc i} $\lambda$7774 and Ca\,{\sc ii} IR are observed (see Figures \ref{fig:qu} and \ref{fig:Pd_Po}), with the O\,{\sc i} feature being slightly more prominent and pointing to a direction opposite to that of Ca\,{\sc ii}. This suggests that the two line-forming regions are inhomogeneously mixed with the rest of the ejecta and spatially distinct from each other. At +5 days the O\,{\sc i} loop has weakened --but is still present-- and the Ca\,{\sc ii} loop has strengthened. By +18 days the O\,{\sc i} loop has completely disappeared and the Ca\,{\sc ii} feature has strengthened even further. This trend could  be explained by the calcium layer being distributed deeper into the ejecta than the oxygen layer, which is consistent with the fact that we found Ca\,{\sc ii} to have a lower velocity than O\,{\sc i} (see Section \ref{sec:lin.specpol}). By +33 days the Ca\,{\sc ii} loop has also started weakening but is still visible, and from +43 days the data are dominated by noise (see Figures \ref{fig:flu_n_pol}, \ref{fig:qu} and \ref{fig:Pd_Po}). 

\begin{table*}
\centering
\caption{\label{tab:comp_table} Maximum degree of polarisation observed in SN~2014ad compared to other broad-lined Type Ic and normal Ib/c SNe.$^*$Unless otherwise stated the dates are given with respect to V-band maximum.$^{\dagger}$ Uncorrected for ISP.}
\begin{tabular}{l l c c c c}
\hline\hline
Type & Name & Date$^*$ & p$_{\text{max}}$(\%) & $\lambda_{\text{max}}^{\dagger}$ (\ang) & Reference\\
\hline
SNe Ic-bl & SN~2014ad & -2 days & 1.62 $\pm0.12$ & 6925 & Section \ref{sec:lin.specpol}\\
SNe Ic-bl & SN~2003dh & +34 d after GRB & 2 $\pm0.5$ & 6500-7500 & \cite{kawabata03}\\
SNe Ic-bl & SN~2006aj & +9.6 days & 3.8 $\pm0.5$ & \about 4200 & \cite{maund06aj}\\
SNe Ic-bl & SN~1998bw & -9 days & 1.1 & 6000-6200 & \cite{patat01}\\
Ic & SN~1997X & + 11 days & 7 & 5000-6000 & \cite{wang01}\\
Ib/c & SN~2005bf & 16 days after first V-band maximum & 4.5 & \about 3800 & \cite{maund05bf}\\
Ic & SN~2008D & + 15 days & 3 $\pm0.5$ & \about 8500 & \cite{maund08D}\\

\hline\hline
\end{tabular}
\end{table*}

\begin{figure}
\centering
\includegraphics[width=8.5cm]{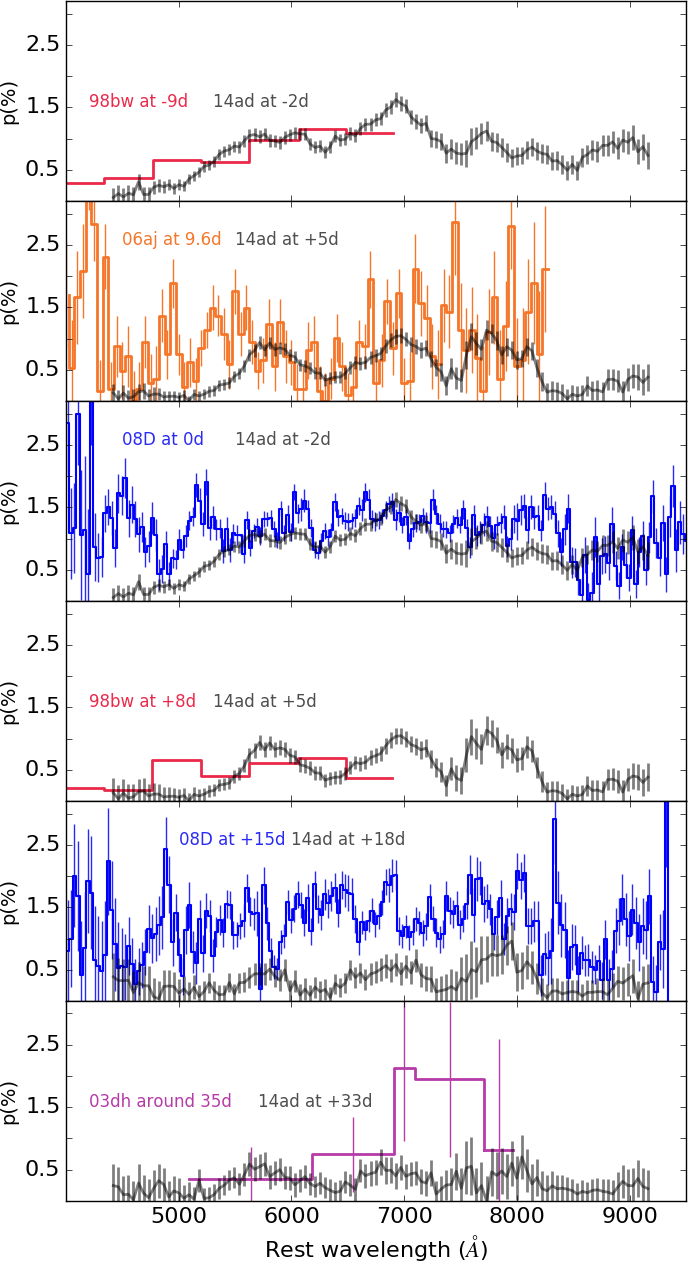}
\caption{\label{fig:comp_pol} Polarisation data (uncorrected for ISP) of SN~1998bw (GRB/Ic-bl), SN~2003dh (GRB/Ic-bl), SN~2006aj (XRF/Ic) and SN~2008D (Ib) compared to the data of SN~2014ad binned to 45 \ang at similar epochs. All epochs are given with respect to V-band maximum, except SN~2003dh which is quoted with respect to the GRB detection date. The data of SN 1998bw were binned to have errors $<$~0.1\% \citep{patat01}.}
\end{figure}

Over the past 20 years spectropolarimetric measurements have been obtained for a number of Ib/c, and SNe Ic-bl with and without GRBs. The maximum level of linear polarisation detected for SN~2014ad is not extremely high compared to previous examples and is similar to SN~1998bw (see Table \ref{tab:comp_table}); it is also the only SN~for which constraints on the circular polarisation have been established. The spectropolarimetric data of SN~2014ad is, to our knowledge, the best data set obtained for a SN~of this type, both in its low level of noise and its large time coverage. For comparison, the data of SN~1998bw (GRB/Ic-bl), SN~2003dh (GRB/Ic-bl), SN~2006aj (XRF/Ic) and SN~2008D (Ib) were plotted along our data at similar epochs, see Figure \ref{fig:comp_pol}. Comparison of the polarisation data of SN~2014ad to that of other SNe can be made difficult by the high levels of noise (e.g. SN~2006aj), broad bins (e.g. SN~1998bw and SN~2003dh) or large error bars (e.g. SN~2003dh) present in their data.

As seen in Figure \ref{fig:comp_pol}, the polarisation of SN~1998bw before maximum follows a trend similar to that seen in our data for SN~2014ad, showing a global increase in polarisation from 4000~\ang\, to 6000~\ang. On the other hand, the data sets of SN~1998bw at +8 days and SN~2014 at +5 days show little resemblance, apart from a similar level of median polarisation over the range 4000~\ang\, to 7000~\ang. In particular, there is a strong discrepancy around 5000~\ang\, where SN~1998bw shows a significant (\about 0.75\%) level of polarisation, whereas SN~2014ad exhibits polarisation that is close to zero. The degree of polarisation observed in the data of SN~1998bw at these wavelengths goes against the assumption that line blanketing due to iron lines in this region causes near complete depolarisation. In the case of SN 2014ad, however, we are confident that this assumption was justified, since the ISP values derived from it were consistent with ISP estimates calculated under the assumption of complete depolarisation at late times (see Section \ref{sec:isp}).

Another comparison of interest is that of the temporal evolution of the degree of polarisation of SN~2014ad and the normal type Ib SN~2008D (see \citealt{maund08D} and references therein). In Figure \ref{fig:comp_pol}, we see that around V-band maximum both SNe show similar levels of polarisation. By +15 days, the level of polarisation of SN~2008D has remained approximately constant whereas that of SN~2014ad has significantly decreased. This indicates that the overall geometry of the photosphere of SN~2008D remains the same closer to the core, whereas SN~2014ad shows evidence of more spherical ejecta in its interior.

When plotted on $q-u$ planes, the spectropolarimetric data of SN~2005bf (Ib/c), SN~2006aj (Ic-bl/XRF), SN~2008D (Ib) and SN~2002ap (Ic-bl) show more scatter and less well defined dominant axes than seen in SN~2014ad at -2 days and +5 days (\citealt{wang03,maund05bf,maund06aj,maund08D}). Additionally, an O\,{\sc i} $\lambda$7774 loop is present in the data of SN~2002ap. Its deviation from the rest of the data is very strong at -6 days and -2 days but subsequently weakens as Ca\,{\sc ii} IR features emerge. By 3 days after V-band maximum the O\,{\sc i} loop in SN~2002ap has nearly disappeared. This evolution is akin to that of the O\,{\sc i} $\lambda$7774 feature in SN~2014ad and our interpretation that the calcium layer is deeper than the oxygen layer is consistent with \cite{wang03}. This suggests that the calcium seen in the data is not primordial but a product of stellar nucleosynthesis and that the ``onion" structure of the stellar interior is partially maintained. On the other hand we also see significant disruption of this structure since the oxygen and calcium line-forming regions were found to be spatially distinct from each other.

As a whole, the spectropolarimetric data reveal ejecta with significant axial symmetry that remain stable for more than 2 weeks after maximum light, and a more spherical interior uncovered by +43 days, whereas one might expect stronger axial symmetry towards the core in the case of a jet driven explosion. Compositional asymmetries are also present, resulting from the partial disruption of the interior structure of the progenitor. Whether these characteristics are consistent with a GRB driven explosion would require modelling, which is beyond the scope of this paper.

\subsection{Spectral modelling and photospheric velocity}
\label{sec:phot_vel}
In order to find values of the photospheric velocity and check our line identification, we created synthetic spectra with \textsc{syn++} (see Section \ref{sec:syn++}). Best results were obtained when using Fe\,{\sc ii}, Na\,{\sc i}, Si\,{\sc ii}, O\,{\sc i} and Ca\,{\sc ii}, but we also tried to fit our data using additional ions. One of our attempts included magnesium, and although it helped suppress the \about 7100~\ang\, peak, it also weakened the Ca\,{\sc ii} emission making the fit to SN~2014ad less accurate. We also tentatively added helium to the synthetic spectrum, in hopes that it may help fit the double dip between the iron and silicon emission. This approach was unsuccessful, and we believe that sodium is the most likely cause of that double dip, although in our fit (shown in Figure \ref{fig:syn++}) the Na\,{\sc i} peak is slightly blue-shifted compared to our data. This may be due to the limitations of a one-dimensional code, which does not account for asphericities in the ejecta.  

The photospheric velocities of SN~2014ad were obtained from model fitting and ranged from -30,000 $\pm$ 5,000~\kms at -2 days to -10,000 $\pm$ 2,000~\kms at 66 days (see Figure \ref{fig:phot_vel}). \cite{modjaz16} studied the spectral properties of 17 Type Ic SN, 10 Ic-bl without observed GRBs and 11 Ic-bl with GRBs, and reported their Fe\,{\sc ii} $\lambda$5169 absorption velocities (used as proxy for the photospheric velocity), see their Figure 5. They also showed that Ic-bl in a GRB/SN~pair tend to have ejecta velocities \about~6000~\kms greater than for SNe Ic-bl without GRB counterparts. Our values of the photospheric velocity for SN~2014ad fit within the regime of the Ic-bl associated with GRBs. With a velocity reaching -30,000 $\pm$ 5,000~\kms at -2 days it even surpasses the average photospheric velocity of SNe Ic-bl with GRBs at the same epoch by approximately -10,000~\kms. Additionally, it is very similar to the expansion velocity observed in SN~1998bw at maximum light \citep{iwamoto98}. The ejecta velocities observed in SN~2014ad are therefore consistent with SNe that have been associated with GRBs.

Irrespective of the way one measures velocities from the flux spectrum (e.g. line fitting, absorption line minimum), the calculated values correspond to the projected velocities, which may differ from velocities depending on viewing angle. In order to put constraints on the estimates made with the 1D code we used to calculate the photospheric velocity, we take as a limiting case that of a spheroid with axis ratio (0.72) given by the highest polarisation recorded for SN 2014ad in Section \ref{sec:lin.specpol} (1.62 \%; \citealt{hoflich91}), and calculate the velocity for a region of the photosphere at the pole of the oblate spheroid. The maximum photospheric velocity obtained at -2 days was -21 750 $\pm$ 3600~\kms, and decreased to -13 775 $\pm$ 3600~\kms by +5 days, which are still within the SNe Ic-bl with GRB regime. 

\textsc{syn++} assumes an infinitely narrow photosphere which does not overlap with the line forming regions, and therefore does not account for electron scattering, as mentioned in section \ref{sec:syn++}. High levels of polarisation are however detected at early days, which indicate the significant role played by electron scattering in the formation of the spectrum. Consequently, our interpretation of just the flux spectrum using  \textsc{syn++} may be incomplete.

\subsection{Metallicity}
\label{sec:met}

\cite{modjaz08}, \cite{levesquekewleyberger10} and \cite{graham15} -- among others -- have compared the metallicity of the host galaxies of Ic-bl with and without GRBs. It has been shown that SNe Ic-bl with GRBs tend to arise in galaxies with lower metallicity than Ic-bl without GRBs, but that the two populations somewhat overlap.

\begin{figure}
\centering
\includegraphics[width=8.5cm]{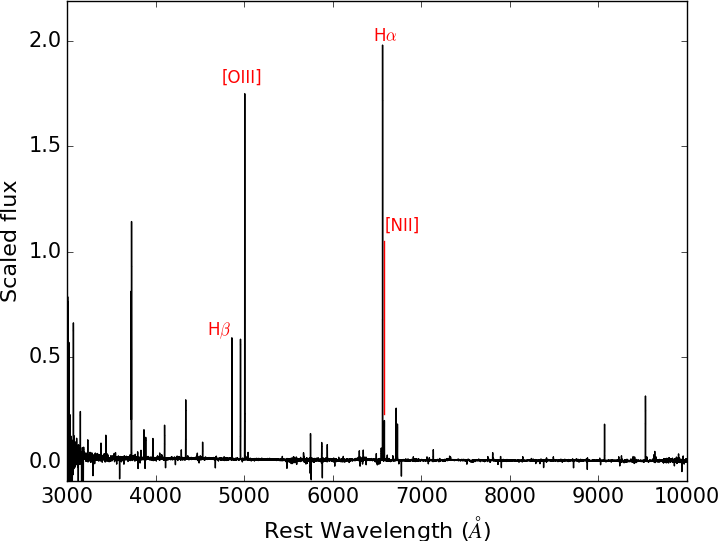}
\caption{\label{fig:xshoot} X-shooter spectrum of the host environment of SN~2014ad obtained on 22 May 2015, 424 days after V-band maximum. The features identified correspond to the lines used for metallicity calculations. }
\end{figure}

\begin{figure}
\centering
\includegraphics[width=8.5cm]{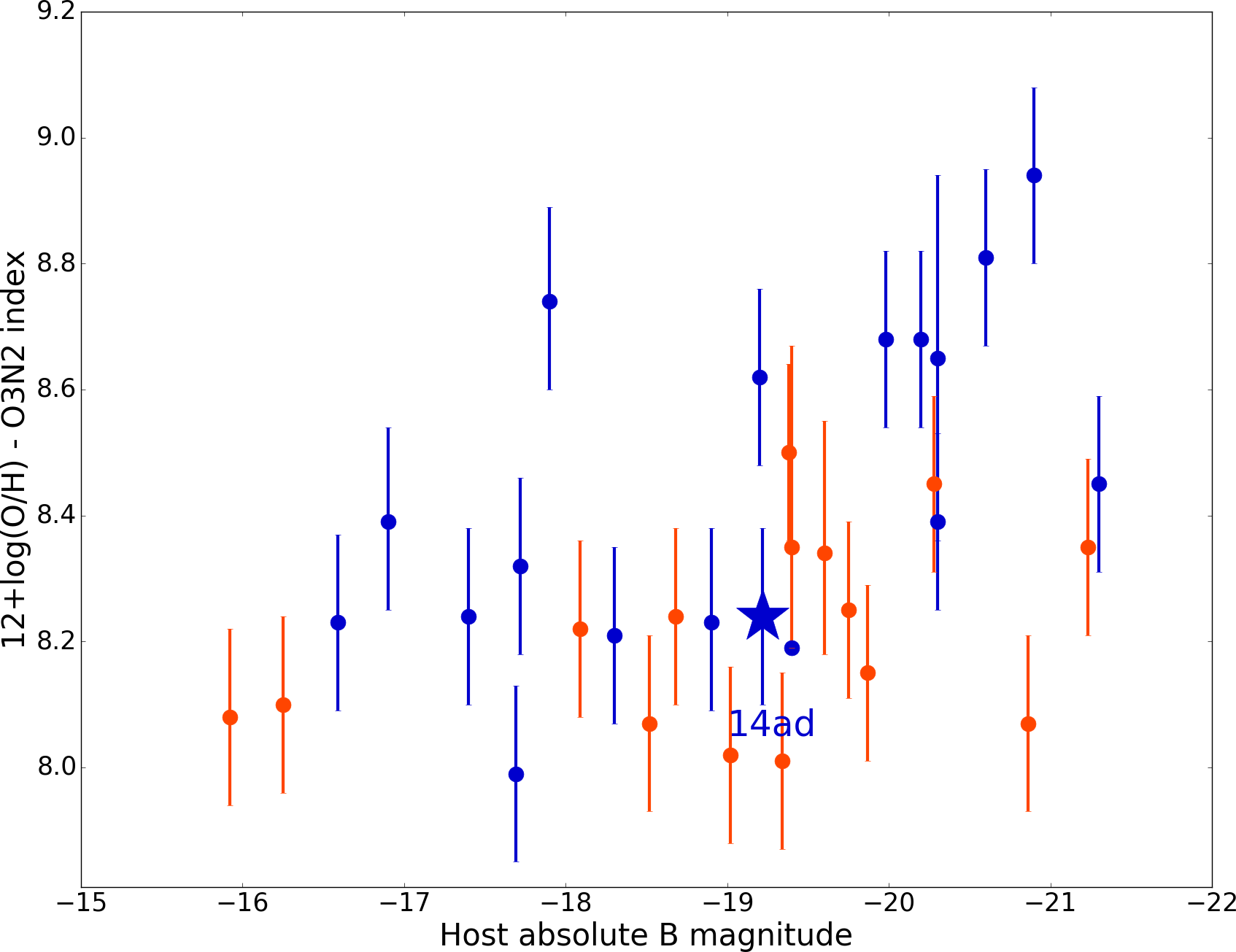}
\caption{\label{fig:met} Comparison of oxygen abundance with host B-band absolute magnitude for 18 SNe Ic-bl (blue) and 15 GRBs (orange). SN 2014ad is represented by a star marker. }
\end{figure}

We compared the metallicity of the host environment of SN~2014ad to that of a sample of SNe Ic-bl -- with and without GRB -- found in the literature. We used the $O3N2$ diagnostic described by \cite{pp04}, and derived the required line fluxes from public spectroscopic X-shooter observations of SN~2014ad obtained on 22 May 2015 \footnote[7]{Under programme 095.D-0608(A). PI: J.Sollerman}, or 424 days after V-band maximum, in which no SN~features are visible. The exposure time was 900 seconds for the blue arm (2970~\ang\, to 5528~\ang) and 960 seconds for the visible arm (5306~\ang\, to 10140~\ang). The data were reduced using the X-shooter pipeline, and the combined spectrum can be seen in Figure \ref{fig:xshoot}. The derived line fluxes are reported in Table \ref{tab:met}. We calculated the  metallicity of the host of SN~2014ad using our own routine and recomputed oxygen abundances of the comparison objects from values of the line fluxes reported in previous works for the \hbeta, [O\,{\sc iii}] $\lambda$ 5007, \halpha and [N\,{\sc ii}]~$\lambda$~6584 lines (see \citealt{modjaz08,graham15}; and references therein). The line fluxes, calculated metallicities and absolute B-band magnitudes for the host galaxies of our comparison objects and SN~2014ad are given in Table \ref{tab:met}. Additionally, we plotted our values of the oxygen abundance against the absolute B-band magnitudes, see Figure \ref{fig:met}.

The oxygen abundance of the host environment of SN~2014ad was found to be \metal = 8.24 $\pm0.14$, which is comparable to both populations of Ic-bl with and without GRBs (see Table \ref{tab:met} and Figure \ref{fig:met}).

\subsection{Late time [O\,{\sc i}] line profile}

\begin{figure}
\centering
\includegraphics[width=8.5cm]{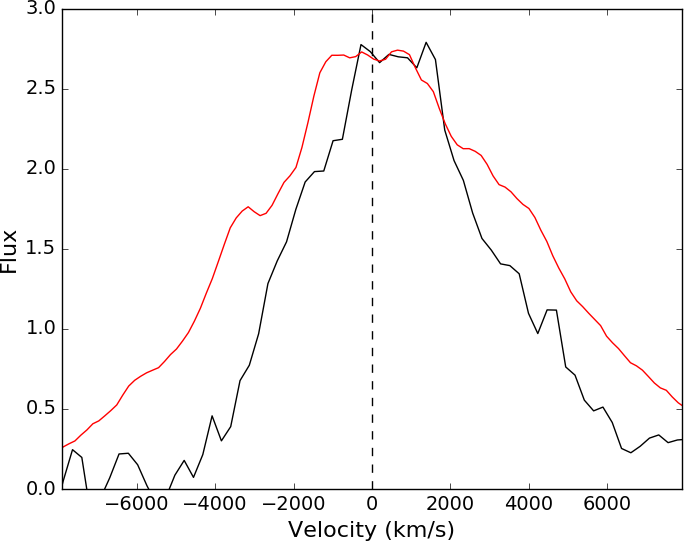}
\caption{\label{fig:OI} Comparison of the [O\,{\sc i}] $\lambda\lambda$6300,6364 line profile of SN~2014ad (red) at 107 days post V-band maximum to SN~1997dq (black) +262 days after estimated outburst \citep{mazzali04}. }
\end{figure}

The shape of the [O\,{\sc i}] line varies significantly from one SN~to another, ranging from single peaks to symmetric double peaks and asymmetric double peaks \citep{maeda08,milisavljevic10}. In the case of SN~2014ad, the [O\,{\sc i}] emission at +107 days mostly resembles a flat-topped (from -1000~\kms to +1000~\kms)  single peak profile, as seen in SN~1997dq (see Figure \ref{fig:OI}; \citealt{mazzali04}). Additionally, a secondary peak is noticeable at -3200~\kms.

The flat top is characteristic of a spherical shell type structure (e.g. \citealt{ignace00, ignace06, maeda08}), however the presence of a blue-shifted feature also indicates contribution from asymmetrically distributed oxygen \citep{milisavljevic10}. Given the low velocity and mostly spherical characteristic of this oxygen line-forming region, it must be separate from the line-forming region responsible for the high velocity asymmetric O\,{\sc i} $\lambda$7774 seen at early days.

Additionally we used the [O\,{\sc i}] line profile as a template to fit the Mg\,{\sc i}] $\lambda$4571 and the [Ca\,{\sc ii}] $\lambda\lambda$7291,7824 lines also present in the spectrum at +107 days (see Figure \ref{fig:spctr}). We found that the FWHM of the feature centered on the Mg\,{\sc i}] $\lambda$4571 line is \about 13,000 km/s, a factor of 2 broader than the feature we identify as [O\,{\sc i}]. Either this is due to blending of lines or a significantly different distribution within the ejecta of the supernova. In other supernovae (e.g. \citealt{spyromilio94}) these features have had very similar profiles. The red side of the calcium forbidden line is well fitted by the [O\,{\sc i}] profile, however the blue side is broader, also suggesting blending with other spectral features. 

\subsection{Toy model of SN 2014ad}

\begin{figure}
\centering
\includegraphics[width=8.5cm]{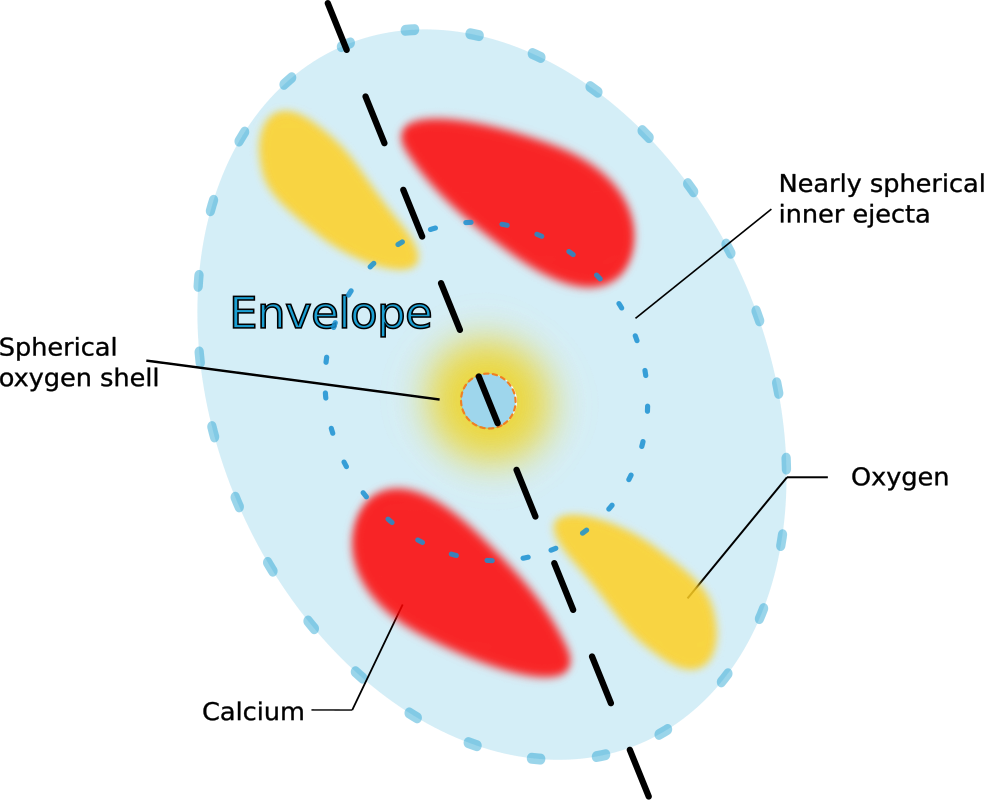}
\caption{\label{fig:toymodel} Toy model of the ejecta of SN~2014ad. The dashed black line represents the dominant axis. Note that the range of the P.A. (0\degree to 180\degree) results in an artificial degree of symmetry.}
\end{figure}

From our interpretation of the spectropolarimetric and spectroscopic data of SN~2014ad we produced a qualitative toy model of the structure of the ejecta (see Figure \ref{fig:toymodel}). The dominant axis present in the Stokes parameters at early days suggests a significant departure from spherical symmetry, but with axial symmetry in the geometry of the outer ejecta. This could translate as an ellipsoidal envelope, the minimum axis ratio of which (0.72) can be constrained by the maximum degree of polarisation we recorded at the first epoch (1.62\%). Departures from axial symmetry were observed in the behaviour of oxygen and calcium suggesting that these line-forming regions are made of large clumps whose angular distribution changes with radius. Additionally the behaviour of the oxygen and calcium features in the  $q-u$ planes revealed that the line-forming regions for the two species must be distinct, and their temporal evolution indicated that the oxygen clumps have higher velocities and dissipate before those of calcium. The decrease in the overall degree of polarisation over time and the disappearance of polarisation features at later epochs suggest that the asymmetries of the outer envelope are not replicated deeper in the ejecta. It is important to note that since the Stokes parameters $q$ and $u$ are quasi-vectors (i.e.. the P.A. ranges from 0\degree to 180\degree) the sketch exhibits an artificial degree of symmetry that may not be representative of the actual structure of the ejecta.

Jet induced models have shown that the distribution of oxygen and calcium tends to be limited to an equatorial torus \citep{khokhlov99}, however in SN~2014ad the line-forming regions are distinct, which is inconsistent with these models. Additionally the Fe blends in the early spectra SN~2014ad correspond to near zero levels of polarisation; \cite{maund05bf} associated the lack of iron polarisation with the absence of jets. Furthermore, \cite{maund05bf} used helium as a tracer for the asymmetric nickel distributions that excite it, which could be caused by a stalled jet in the envelope; unfortunately the lack of helium in SN~2014ad denies us any information about the nickel. The decreasing level of polarisation over time is also difficult to reconcile with the ``choked" jet model proposed for SN~2005bf and SN~2008D (\citealt{maund05bf,maund08D}). Those conclusions, however, were drawn on limited observations of those SNe and serve to highlight the importance of multi-epoch spectropolarimetry for achieving truly 3D tomography of SNe.

\section{CONCLUSION}
We report 7 epochs of spectropolarimetry for the Ic-bl SN~2014ad ranging from -2 days to + 66 days, as well as 8 epochs of spectroscopy from -2 days to +107 days. This is to our knowledge the best spectropolarimetric data set available for a Ic-bl SN. A maximum degree of polarisation of 1.62 $\pm$0.12 \% was detected at -2 days, and the level of ISP was found to be 0.15 $\pm$0.06\%. 
A clear dominant axis in the $q-u$ plots corresponding to the first 3 epochs was found with a P.A. \about 20\degree, indicating significant axial symmetry. The decrease in the overall level of polarisation over time and the disappearance of the dominant axis at +33 days suggests that the inner ejecta are more spherical than the outer layers. 

Polarisation features associated with O\,{\sc i} and Ca\,{\sc ii} show departures from the dominant axis in the form of loops, suggesting line-forming regions in the form of large clumps that are asymmetrically distributed. Since the oxygen and calcium loops are oriented in different directions the two line-forming regions must be distinct from each other, but both are moving very rapidly. Additionally, due to the separate temporal evolution of their polarisation signatures (oxygen appears and disappears earlier than calcium) and different velocities (oxygen is faster than calcium)  we conclude that the calcium layer is deeper into the ejecta than this oxygen layer. This means that the calcium we observe is not primordial but a product of stellar nucleosynthesis and that the ``onion" structure of the progenitor is at least partially preserved. 

At +107 days the [O\,{\sc i}] line profile suggests the presence of a deeper oxygen line-forming region in a near spherical shell, unlike the oxygen observed at earlier dates. This is consistent with the decreasing levels of polarisation detected deeper in the ejecta. 

Using X-shooter data at +424 days we calculated the metallicity of the host galaxy of SN~2014ad and found it was rather low and consistent with the population of Ic-bl with GRB counterparts. It should be noted that it is also close to the metallicity of the host galaxy of two other Ic-bl without GRBs with similar B-band magnitude: SN~2006nx and SN~2007qw. Because of the overlap in the two populations, the low metallicity of the host environment of SN~2014ad is inconclusive. 

It is not clear from the spectropolarimetric data of SN~2014ad whether the explosion was driven by jets: the axial symmetry at early epochs is not inconsistent with such scenario, but the more spherical inner ejecta is more difficult to reconcile with the presence of a jet. The high velocity oxygen and calcium line-forming regions follow directions that are similar to that of the dominant axis, which would correlate with the direction of jets, should they be present. Hydrodynamical simulations of asymmetrical explosions produced by \cite{maeda02} to model SN 1998bw showed low-velocity oxygen being ejected in the equatorial plane, which is inconsistent with our picture of SN 2014ad. On the other hand, fully jet-driven explosion models by \cite{couch11} revealed intermediate mass elements being entrained by the outflow and as a result roughly following the direction of the jets, which is consistent with the behaviour of the high velocity calcium and oxygen in SN 2014ad. The potential outflow may not have fully broken out of the envelope and produced a GRB, but could have been sufficiently powerful to accelerate the ejecta as fast as 30,000 $\pm$ 5,000~\kms and create the strong alignment observed in the $q-u$ planes until 18 days after V-band maximum. This scenario is consistent with the conclusions of \cite{lazzati12}, who showed that for an engine of short enough life time  the jets may dissipate through the envelope before reaching the surface, resulting in a relativistic SN~explosion but no GRB. It is, however, difficult to reconcile the case of ``choked" jets with nearly spherical ejecta, as observed in SN~2014ad. If SN~2014ad did have a GRB counterpart, it may have been collimated away from our line of sight, and without radio observation we cannot exclude or support this scenario with complete certainty.

\section*{Acknowledgements} 
The authors would like to thank the staff of the Paranal Observatory for their kind support and for the acquisition of such high quality data on the program 093.D-0820. We are also very grateful to Emma Reilly for triggering the observations. HFS is supported through a PhD scholarship granted by the University of Sheffield. The research of JRM is supported through a Royal Society University Research Fellowship. JCW is supported by NSF Grant AST 11-09881.

\bibliographystyle{mn2e}

\appendix

\section{Oxygen abundance comparison}

\begin{table*}
\centering
\begin{tabular}{l c c c c c c c }
\hline \hline 
Ic-bl and GRB Hosts & \hbeta & [OIII] $\lambda$5007 & \halpha & [NII] $\lambda$6584 & \metal & M$_B$ & Spectrum
\\\hline 
SN1997ef$^*$ & 478 $\pm$ 51 & 223 $\pm$ 30 & 1730 $\pm$ 170 & 581 $\pm$ 62 & 8.68$\pm$ 0.14 & -20.2$^1$ & Flux$^1$ \\
SN2003jd$^*$ & 229 $\pm$ 24 & 342 $\pm$ 36 & 752 $\pm$ 76 & 98 $\pm$ 11 & 8.39 $\pm$ 0.14 & -20.3$^1$ & Flux$^1$ \\
SN2005kz$^*$ & 656 $\pm$ 12 & 115 $\pm$ 17 & 1660 $\pm$ 170 & 1300 $\pm$ 130 & 8.94 $\pm$ 0.14 &-20.9$^1$ & Flux$^1$ \\
SN2005nb$^*$ & 376 $\pm$ 45 & 455 $\pm$ 52 & 1510 $\pm$ 150 & 299 $\pm$ 33 & 8.45$\pm$ 0.14 & -21.3$^1$ & Flux$^1$ \\
SN2005kr & 8.20 $\pm$ 0.83 & 24.3 $\pm$ 2.4 & 28.4 $\pm$ 2.8 & 2.41 $\pm$ 0.26 & 8.24 $\pm$ 0.14 & -17.4$^1$ & Flux$^1$ \\
SN2005ks &  69.9 $\pm$ 7.3 & 51.2 $\pm$ 5.5 & 272 $\pm$ 27 & 92.3 $\pm$ 9.3 & 8.62 $\pm$ 0.14 & -19.2$^1$ & Flux$^1$ \\
SN2006nx & 6.9 $\pm$ 1.1 & 20.9 $\pm$ 2.2 & 33.1 $\pm$ 3.4 & 2.73 $\pm$ 0.51 & 8.23 $\substack{+0.15 \\ -0.14}$ & -18.9$^1$ & Flux$^1$ \\
SN2006qk & 43.7 $\pm$ 5.0 & 15.3 $\pm$ 2.6 & 230 $\pm$ 23 & 86.2 $\pm$ 8.8 & 8.74 $\substack{+0.15 \\ -0.14}$ & -17.9$^1$ & Flux$^1$ \\
SN2007I & 28.7 $\pm$ 2.9 & 57.2 $\pm$ 7.1 & 119 $\pm$ 13 & 20.0 $\pm$ 3.2 & 8.39 $\substack{+0.15 \\ -0.14}$ & -16.9$^1$ & Flux$^1$ \\
SN2002ap & 1.285 & 0.212 & 6.007 & 1.812 & 8.81 $\pm$ 0.14 & -20.6$^1$ & Hbeta$^2$\\
SN2007ce & 108.4 & 594.8 & 285.1 & 7.813 & 7.99 $\pm$ 0.14 & -17.69$^3$ & Hbeta$^4$\\
SN2008iu & 117.1 & 745.2 & 255.1 & 43.25 & 8.23 $\pm$ 0.14 & -16.59$^3$ & Hbeta$^4$\\
SN2009bb & 37.36 & 17.56 & 155.3 & 51.59 & 8.68 $\pm$ 0.14 & -19.98$^5$ & Flux$^5$\\
SN2010ah & 104.0 & 186.9 & 359.3 & 33.88 & 8.32 $\pm$ 0.14 & -17.22$^3$ & Hbeta$^4$\\
SN2010ay & 271.6 & 905.0 & 837.2 & 67.20 & 8.21 $\pm$ 0.14 &  -18.30$^6$ & Flux$^3$\\
SN2014ad & 9.35 $\pm$ 0.33 & 28.53 $\pm$ 0.38 & 33.25 $\pm$ 0.14 & 3.09 $\pm$ 0.11 & 8.24 $\pm0.14$ & -19.22 & Flux\\
GRB980425/1998bw & 219.69 & 851.4 & 713.7 & 68.7 & 8.22 $\pm$ 0.14 & -18.09$^1$ & Hbeta$^1$\\
GRB991208/... & 3.493 & 5.848 & 17.04 & 0.852 & 8.24 $\pm$ 0.14 & -18.68$^1$ & Flux$^7$\\
GRB010921/... & 9.518 & 29.65 & 40.11 & 1.858 & 8.15 $\pm$ 0.14 & -19.87$^1$ & Flux$^{7,8}$\\
GRB011121/... & 17.13 & 16.64 & 65.61 & 2.0 & 8.25 $\pm$ 0.14 & -19.75$^1$ & Flux$^9$\\
GRB020819B/Dark & 3.0 $\pm$ 0.9 & 9.7 $\pm$0.7 & 14.5 $\pm$ 1.2 & 2.7$\pm$1.7 &  8.34 $\substack{+0.21\\ -0.16}$& -19.6$^1$ & Flux$^{10}$\\
GRB020903/unnamed & 44.0 & 335.0 & 168.0 & 7.2 & 8.01 $\pm$ 0.14 & -19.34$^1$ & Flux$^{11}$\\
GRB030329/2003dh & 1 & 3.40 & 2.74 & 0.1 & 8.10 $\pm$ 0.14 & -16.52$^1$ & Hbeta$^{7,12}$\\
GRB031203/2003lw & 1 & 6.36 & 2.82 & 0.15 & 8.07 $\pm$ 0.14 & -18.52$^1$ & Hbeta$^{12}$\\
GRB050824/... & 2.529 & 15.45 & 7.6 & 0.2797 & 8.02 $\pm$ 0.14 & -19.02$^1$ & Flux$^{13}$\\
GRB050826/... & 28.65 & 36.22 & 85.22 & 14.41 & 8.45 $\pm$ 0.14 & -20.28$^1$ & Flux$^7$\\
GRB051022/Dark & 25.29 & 59.57 & 104.99 & 15.97 & 8.35 $\pm$ 0.14 & -21.23$^1$ & EQW$^{3}$\\
GRB060218/2006aj & 68.83 & 229.9 & 170.5 & 5.122 & 8.08 $\pm$ 0.14 & -15.92$^1$ & Flux$^{14}$\\
GRB060505/Dark & 4.553 & 5.500 & 22.85 & 5.185 & 8.50 $\pm$ 0.14 & -19.38$^{3,15}$ & Flux$^{15}$\\
GRB070612A/... & 36.60 & 41.01 & 152.0 & 1.520 & 8.07 $\pm$ 0.14 & -20.86$^3$ & Flux$^7$\\
GRB120422A/SN2012bz$^*$ & 0.5 $\pm$ 0.4 & 1.9 $\pm$ 0.2 & 2.4 $\pm$ 0.1 & 0.6 $\pm$ 0.2 & 8.35$\substack{+0.32 \\ -0.16}$ & - 19.4$^{16}$ & Flux$^{16}$ \\ \hline 
SN2002bl$^*$ & ... & ... & ... & ... & 8.65 $\pm$ 0.3$^{17}$ & -20.3$^{17}$ & ...\\
SN2007qw$^*$ & ... & ... & ... & ...  & 8.19 $\pm$ 0.01$^{17}$& -19.4$^{17}$ & ...\\
\hline \\

\end{tabular}
\caption{\label{tab:met} Line fluxes, O3N2 metallicity and B-band absolute magnitude of the host environments of BL Type Ic SNe and GRBs. The SNe associated with the GRBs are given when known, "Dark" indicates the absence of an optical counterpart, and an ellipse "..." is shown when a SN~cannot be ruled out because the searches conducted were not deep enough. The emission line measurements correspond to the values for the centre of the host galaxy, unless the object is marked with an asterisk ($^*$), in which case they coincide with the host environment of the object within the host galaxy. The line fluxes and equivalent widths are all in units of 10$^{-17}$ erg s$^{-1}$ cm$^{-2}$. When the values found of the literature did not have associated errors, we assumed an uncertainty \about 10\%. The last column indicates whether the values for a given object correspond to the flux values, the equivalent width, or the flux value normalised with respect to \hbeta (Note that the \hbeta column of an \hbeta normalised flux might not be 1 or 100 if the galactic extinction correction was applied to normalised values in the literature). The metallicity of SN~2002bl and SN~2007qw are taken directly from (17). \textbf{References:} (1) \cite{modjaz08}; (2) \cite{ferguson98}; (3) \cite{graham13}; (4) \cite{sanders12}; (5) \cite{levesquesoderberg10}; (6) SDSS-mpg; (7) \cite{levesquekewleyberger10}; (8) \cite{price12}; (9) \cite{garnavich03}; (10) \cite{perley17}; (11) \cite{hammer06}; (12) \cite{sollerman05}; (13) \cite{mcglynn07}; (14) \cite{levesquebergerkewley10}; (15) \cite{thone08}; (16) \cite{schulze14}; (17) \cite{modjaz11}. }
\end{table*}

We compared the oxygen abundance of the host galaxy of SN~2014ad to a sample of 32 SNe and GRBs listed in Table \ref{tab:met}. They include three dark GRBs and 6 bursts for which the presence of a SN~in the afterglow could not be ruled out. It is crucial to compare metallicities computed with the same diagnostic as different metallicity indicators yield slightly different values. We re-computed the oxygen abundance for all but two (SN~2002bl and SN~2007qw) of our comparison SNe as we could not find values for the line fluxes of \hbeta [O\,{\sc iii}] $\lambda$5007, \halpha and [N\,{\sc ii}] $\lambda$6584. We used our own routines and line fluxes taken from the literature. Unfortunately a lot of the flux values reported did not have errors. We assumed uncertainties of the order 10\% of the nominal value of the flux, which is consistent with \cite{modjaz08}. It should be noted, however, that the main source of error when calculating the $O3N2$ metallicity are the systematic errors (0.14 dex) of this particular diagnostic \citep{pp04}. The line fluxes and derived oxygen abundances are summarised in Table \ref{tab:met}.

\end{document}